\documentclass{aa}
\usepackage[varg]{txfonts}
\usepackage{graphicx}
\usepackage{times}
\usepackage{color}
\usepackage{gensymb}
\usepackage{placeins}
\usepackage{afterpage}

\definecolor{customhdrcolor}{rgb}{0.0,0.0,0.0}
\definecolor{customcitecolor}{rgb}{0.0,0.5,0.75}
\definecolor{customlinkcolor}{rgb}{0.0,0.5,0.75}

\usepackage[colorlinks=true,linkcolor=customlinkcolor,urlcolor=customlinkcolor,citecolor=customcitecolor,pdftex]{hyperref}
\usepackage[caption=false]{subfig}

\ifpdf\else\usepackage{graphics}\fi

\newcommand\Tstrut{\rule{0pt}{2.6ex}}         
\newcommand\Bstrut{\rule[-0.9ex]{0pt}{0pt}}   

\DeclareRobustCommand{\TUSSEN}[3]{#2}

\begin{document}

\title{Precision requirements for interferometric gridding\\ in 21-cm power spectrum analysis}


\author{A.~R.~Offringa\inst{1,2}
	\and F.~Mertens\inst{2}
	\and S.~van~der~Tol\inst{1}
	\and B.~Veenboer\inst{1}
	\and B.~K.~Gehlot\inst{2,3}
	\and L.~V.~E.~Koopmans\inst{2}
	\and M.~Mevius\inst{1}
}

\institute{
	Netherlands Institute for Radio Astronomy (ASTRON), 7991 PD Dwingeloo, The Netherlands. \email{offringa@astron.nl}
	\and Kapteyn Astronomical Institute, University of Groningen, PO Box 800, NL-9700 AV Groningen, the Netherlands
	\and School of Earth and Space Exploration, Arizona State University, 781 Terrace Mall, Tempe, AZ 85287, USA
}

\date{Received 18/04/2019 / Accepted 27/08/2019}

\label{firstpage}
  \abstract
   {Experiments that try to observe the 21-cm redshifted signals from the Epoch of Reionization using interferometric low-frequency instruments have stringent requirements on the processing accuracy. }
   {We analyse the accuracy of radio interferometric gridding of visibilities with the aim to quantify the power spectrum bias caused by gridding, ultimately to determine the suitability of different imaging algorithms and gridding settings for 21-cm power spectrum analysis. }
   {We simulate realistic Low-Frequency Array (LOFAR) data, and construct power spectra with convolutional gridding and $w$-stacking, $w$-projection, image domain gridding and without $w$-correction. These are compared against directly Fourier transformed data. The influence of oversampling, kernel size, $w$-quantization, kernel windowing function and image padding are quantified. The gridding excess power is measured with a foreground subtraction strategy, for which foregrounds have been subtracted using Gaussian progress regression, as well as with a foreground avoidance strategy.}
   {Constructing a power spectrum that has a bias significantly lower compared to the expected EoR signals is possible with the tested methods, but requires a kernel oversampling factor of at least 4000 and, when using $w$-correction, at least 500 $w$-quantization levels. These values are higher than typical values used for imaging, but are computationally feasible. The kernel size and padding factor parameters are less crucial. Among the tested methods, image domain gridding shows the highest accuracy with the lowest imaging time. }
   {LOFAR 21-cm power spectrum results are not affected by gridding. Image domain gridding is overall the most suitable algorithm for 21-cm Epoch of Reionization power spectrum experiments, including for future Square Kilometre Array (SKA) EoR analyses. Nevertheless, convolutional gridding with tuned parameters results in sufficient accuracy for interferometric 21-cm Epoch of Reionization experiments. This holds also for $w$-stacking for wide-field imaging. The $w$-projection algorithm is less suitable because of the kernel oversampling requirements, and a faceting approach is unsuitable due to the resulting spatial discontinuities. }


   \titlerunning{Gridding for 21-cm power spectrum analysis}
   \authorrunning{A. R. Offringa et al.}
   
   \maketitle
   
%

\section{Introduction}
The Epoch of Reionization (EoR) is a pivotal era in the evolution of our Universe. In this era, that is expected to have started approximately 500 million years after the Big Bang, the very first objects in our Universe heated and ionized the intergalactic medium. One of the most promising ways to analyse this process is through detection of the redshifted 21-cm line emission of neutral Hydrogen \citep{liev-2002, morales-2005, furlanetto-2006, mcquinn-2006, pritchard-2012, park-2019}. Current constraints indicate that the EoR has taken place at a redshift of approximately $z=6-10$, implying that the 21-cm signals from the EoR are detectable in the frequency range of approximately $130$--$200$~MHz. Several low-frequency instruments have been built or are planned with the detection of these signals as one of their key science goals \citep{parsons-2012-arraysensitivity, lofar-2013, tingay-2013-mwa, dewdney-2014-ska, deboer-2017-hera, fialkov-2018-LEDA}.

Interferometric experiments aim to detect the EoR signals through power spectrum analysis \citep{paciga-2013-gmrt-eor, beardsley-2016, patil-2017, trott-2016-chips}. Such analysis can combine a large field of view and several MHz of bandwidth to decrease the uncertainty due to the thermal noise of the instrument to ultimately detect the signals from the EoR in a statistical manner. Recently, the LOFAR EoR project is working on interferometric upper limits from 10 nights of observations \citep{mertens-10night-eorlimit-2019}. Direct imaging of the EoR is probably not feasible until the Square Kilometre Array (SKA; \citealt{dewdney-2014-ska}) is functional \citep{zaroubi-2012}.

It is often necessary to average visibility measurements together that observe (almost) the same modes on the sky in order to deal with the enormous volume of visibilities that are produced by modern telescopes. One method to do this is by gridding the visibilities on a regular 2D grid in Fourier ($uv$) space, a step that is also part of making images from interferometric data. This step leads to small errors. Gridding can be avoided in some specific power spectrum methodologies, such as by making use of redundancy \citep{parsons-2009} or by using different transforms based on spherical harmonics or $m$-mode analysis \citep{carozzi-2015,ghosh-2018,eastwood-2018}.

Several 21-cm power spectrum pipelines use gridded $uv$-cubes or image cubes (the third dimension being frequency), such as \textsc{chips} \citep{trott-2016-chips}, a pipeline that constructs a fully invariance-weighted power spectrum; the \textsc{$\epsilon$ppsilon} pipeline \citep{jacobs-2016, barry-2019-eppsilon} that makes use of the Fast Holographic Deconvolution software for imaging (\textsc{fhd}; \citealt{sullivan-fhd-2012}); and the image-based tapered gridded estimater (\textsc{itge}; \citealt{choudhuri-2018}). In this paper we make use of the two LOFAR 21-cm power spectrum pipelines described in \citet{offringa-2019}, which use \textsc{wsclean} \citep{offringa-wsclean-2014} for imaging the data. During calibration or source subtraction, gridded model images are sometimes used to forward predict (continuous) model visibilities. This requires the reverse action of gridding, sometimes referred to as de-gridding, and like gridding, this is also subject to small errors.

In typical scenarios gridding decreases the number of data samples by several orders of magnitude. Besides decreasing the data volume, gridding may also have some other benefits:
\begin{itemize}
 \item By imaging only the most sensitive part of the primary beam, emission that falls outside the imaged area is removed. Sidelobes of off-axis emission are not removed. Off-axis emission is often harder to model and calibrate, and removing this emission can, therefore, be a benefit. In the context of power spectrum analysis, this might come at the cost of no longer being able to measure the largest scales and increasing the sample variance \citep{choudhuri-tge-2016}.
 Alternatively, visibility-based filters exist that allow some degree of primary beam shaping without gridding \citep{post-correlation-filtering, parsons-beamsculpting-2016, atemkeng-2016}, but these are more limited than what is provided by a gridding anti-aliasing filter or by trimming or windowing in image space.
 \item During gridding, projection algorithms for correcting direction-dependent effects can be included, such as the $w$-term, the primary beam and the ionosphere \citep{wprojection-cornwell, bhatnagar-aproj-2008, bhatnagar-2013-wb-aprojection, tasse-2013-awimager}.
 \item The output of the gridding algorithm can be stored as a standardized product (e.g. a \textsc{fits} image cube), which improves the overall modularity of a pipeline, making it easier to analyse and compare with different gridders or power spectrum pipelines. This can help in localizing the cause of excess power (such as foreground sources that have not been subtracted properly) and allows using code from regular imaging software that is rigorously tested.
\end{itemize}

Separating the redshifted 21-cm signals from the Galactic and extra-galactic foregrounds requires a high dynamic range: whereas the foregrounds have a brightness temperature of a few thousand kelvin, the expected 21-cm signals are only a few mK. In order to use an approach that includes gridding, the gridding algorithm needs to have a high accuracy in order not to bias the power spectrum measurements, while it is at the same time necessary to process large data volumes within a reasonable time. In this paper, we analyse the influence of gridding on the accuracy of the 21-cm power spectrum. We will investigate the magnitude at which the power spectrum is affected by the gridding, and analyse the minimal required conditions that makes the power spectrum bias sufficiently small to be able to detect 21-cm signals from the EoR or the Cosmic Dawn with their expected signal strength.

In Section \S\ref{sec:gridding}, we describe gridding methods and list their accuracy trade-offs. Section \S\ref{sec:power-spectra} describes the methodology to calculate power spectra from gridded images. The simulated data is described in Section \S\ref{sec:data}, and the gridding accuracy test results are presented in Section \S\ref{sec:results}. In Section \S\ref{sec:discussion}, we discuss the results and draw conclusions.

\section{Gridding} \label{sec:gridding}
To understand the effects caused by interferometric gridding, we start by describing some of the foundations of gridding. An interferometer samples the complex visibility function
\begin{align}\label{eq:visibility-function}
V(u,v,w) = \iint \frac{A(l,m) I(l,m)}{\sqrt{1-l^2-m^2}} 
e^{-2\pi i \left(ul + vm + w(\sqrt{1-l^2-m^2}-1)\right)} dl dm,
\end{align}
where $u,v,w$ specifies a baseline coordinate in the coordinate system of the array, $A$ is the primary-beam function, $I$ is the sky function and $l,m$ specifies a cosine sky coordinate. The visibility function $V$ is the result of interferometric observing and calibration. In this work, we ignore any errors that might occur during this process.

When doing polarimetry, $V$, $A$ and $I$ become $2\times 2$ matrices. Without loss of generality we will ignore polarisation and treat imaging as a scalar problem. We will also not cover gridding with the element beam and instead assume $A$ to be unity. Application of the element beam during gridding potentially improves the sensitivity of the power spectrum, because this allows including the primary-beam weighted full field of view into the power spectrum. However, the improvement in power spectrum sensitivity of gridding with the beam is small, because most of the sensitivity is achieved by the central part of the beam. Using only the most sensitive part of the beam avoids parts of the beam that are less well modelled, and this has therefore been the LOFAR EoR approach in practice \citep{patil-2017}.

Imaging consists of solving $I$ from $V$, thereby inverting Eq.~\eqref{eq:visibility-function}. Part of imaging consists of calculating the PSF-convolved (dirty) image $I'$,
\begin{equation}
I'(l,m) = \frac{\sqrt{1-l^2-m^2}}{N} \int \mathcal{F_V}(l, m, w) e^{2\pi i w(\sqrt{1-l^2-m^2}-1)} dw
\end{equation}
with $N$ a normalization constant that corrects for the $uv$-coverage and $\mathcal{F_V}$ the inverse 2D Fourier transform of visibilities $V$ with the same $w$-value,
\begin{equation}
\mathcal{F_V}(l, m, w) = \iint V(u,v,w) e^{2\pi i \left(ul + vm\right)} du dv.
\end{equation}
We do not consider deconvolution in this paper. It is common to subtract bright sources before the gridding step \citep{beardsley-2016, patil-2017, trott-2016-chips}.

Gridding consists of discretizing the non-uniformly sampled $u,v,w$ values. We consider gridding with and without $w$-term correction, and investigate the accuracy that different $w$-term correcting methods achieve. The simplest method of gridding is by adding the value of each visibility to the closest $uv$ grid point (nearest-neighbour gridding) and ignoring its $w$-value. Such gridding introduces two types of errors:
\begin{enumerate}
 \item Aliasing: Visibilities and the $uv$-sampling function might have frequencies beyond the corresponding Nyquist-rate of the $uv$-grid (i.e., they are not band-limited at the resolution of the $uv$-grid). In other words, sources and sidelobes might exist outside the imaging field of view. Structures outside the field of view are aliased, appearing as ghost structures within the imaging field of view.
 \item Discretization of $u$, $v$ and $w$-values: The true continuous $uv$-value of the sample is discretized to match the regular $uv$-grid. This causes smearing and decorrelation of emission. Similarly, any non-coplanarity of the array causes visibilities with different $w$-terms to be averaged together, also leading to smearing and decorrelation.
 \end{enumerate}
Visibilities can be band-limited by low-pass filtering the visibilities, thereby avoiding aliasing. The common method to do this is by convolving the visibilities with a smoothing kernel --- the so-called anti-aliasing kernel \citep{brouw-1975, schwab-1980}. Gridding with the element beam $A$ (see Eq.~\ref{eq:visibility-function}) can acts as a natural anti-aliasing kernel \citep{bhatnagar-aproj-2008}. In case the convolution kernel is a continuous function (such as a sinc function), convolutional gridding has the additional benefit that the contribution of the continuous visibilities can be evaluated precisely at each discretized $uv$-position, which solves the second inaccuracy (i.e., point 2 listed above) for $u$ and $v$. The $w$-term can be corrected by one of several $w$-correction methods, such as convolving each visibility with a $w$-correction term that projects it onto the $w=0$ plane \citep{wprojection-cornwell}.

By convolving each visibility with the combination of an anti-alias kernel and a $w$-term correction kernel, it is theoretically possible to perform gridding that matches the accuracy of a direct Fourier transform. In practice, gridders apply further simplifications for various reasons:
\begin{itemize}
 \item To reduce the computational cost of the convolution, the spatial anti-aliasing low-pass kernel is windowed to a typical size of 7 $uv$-cells. The prolate spheroidal wave function is commonly used as a compact low-pass filtering kernel and has several beneficial properties for gridding \citep{brouw-1975}. It is sometimes approximated by the easier to evaluate Kaiser-Bessel function (e.g., \citealt{offringa-wsclean-2014}).
 \item The kernels are precalculated and interpolated to avoid evaluation of a computationally expensive function for each visibility. This requires discretization of the $uv$ space in which it can be evaluated, resulting in errors. It is possible to more finely sample the kernel and thereby reduce the error. We will refer to the factor by which the kernel is increased as the oversampling factor.
 \item Because of the precalculation of kernels, the $w$-values are discretized as well. The number of $w$-discretization levels can strongly affect the computational performance.
 \item To limit the size of the kernel in the case of $w$-projection, the $w$-kernel is trimmed at a point at which its power is a small fraction (e.g. 1\%) of its peak power. This error is not made in a pure $w$-stacking algorithm, since the $w$-term is applied in the full image domain.
\end{itemize}

In this paper, we use \textsc{wsclean} as a gridding and imaging platform, which implements several gridding engines: a $w$-stacking gridder (\S\ref{sec:wstack}); an image-domain gridder (\S\ref{sec:idg}); and inversion using a direct Fourier transform. The latter implements an imaging operation that is computationally the most expensive but accurate up to the floating-point precision, and is used as the ground truth in this work. We use \textsc{wsclean} version~2.6, released on 2018-06-11. \textsc{wsclean} is open-source.\footnote{The \textsc{wsclean} software is available at \url{http://wsclean.sourceforge.net/}.} Even though we use a specific gridder implementation, we analyse generic gridding parameters which are applicable to most imaging algorithms. We include an analysis of standard convolutional gridding by turning $w$-stacking off. These results are therefore applicable to any standard (e.g. prolate-spheroidal based) gridding implementation, such as the implementation in \textsc{casa}. Additionally, because \textsc{wsclean} does not implement $w$-projection, we include an analysis of the $w$-projection implementation in \textsc{casa} \citep{casa-2007}.

\subsection{$w$-stacking gridding} \label{sec:wstack}
In the $w$-stacking algorithm, visibilities are gridded on a number of $w$-planes, each corresponding to a certain range of $w$-values. All planes are separately Fourier transformed to the image domain, and the $w$-term is subsequently corrected for by applying multiplication of the images by the spatially-varying $w$-term. The standard gridding engine of \textsc{wsclean} applies the $w$-stacking algorithm to correct for $w$-terms \citep{offringa-wsclean-2014}. This gridder is used in this work for investigating the influence of gridding settings on the power spectrum.

Configurable gridding parameters that we will investigate are:
\begin{itemize}
 \item Anti-aliasing kernel size --- the width of the convolution kernel in number of uv-cells. The \textsc{wsclean} default for this setting is 7, which indicates that the kernel covers $7 \times 7$ $uv$-cells.
 \item Kernel oversampling factor --- for performance reasons, the kernel is tabulated beforehand and not directly evaluated. When a value is gridded on the $uv$-plane, the nearest tabulated kernel is selected. Other interpolation methods such as linear interpolation help to reduce the error, but increase the per-visibility cost and are not implemented in \textsc{wsclean}. In \textsc{wsclean}, the default is to oversample the kernel 63 times, which implies a precomputed table of size $7 \times 63$.
 \item Gridding function --- By default, \textsc{wsclean} uses a sinc function windowed by a Kaiser-Bessel function \citep{kaiser-1980}, which approximates a discrete prolate spheroidal sequence (DPSS).
 \item Padding factor --- Factor by which the image size is increased beyond the field of interest, to avoid edge issues. By default, \textsc{wsclean} uses a factor of 1.2.
 \item Number of $w$-layers -- Discrete number of $w$-values. Visibilities are moved to their nearest $w$-value. By default, \textsc{wsclean} uses a number of $w$-values such that the maximum phase decorrelation, which occurs at the edge-pixels of the image, is 1 radian.
\end{itemize}
 
In the $w$-stacking implementation, all calculations are performed with 64-bit (IEEE 754-2008) double-precision floating-point values.

\subsection{Image domain gridding} \label{sec:idg}
Image domain gridding (\textsc{idg}; \citealt{vandertol-idg-2018}) is a method that calculates the contribution of visibilities in image space. Visibilities are grouped into slightly overlapping $uv$-subgrids, each covering a small part of the $uv$-plane (typically 32$^2$ to 128$^2$ cells). The contribution of the visibilities in their subgrid is then calculated by evaluating the image-domain ($lm$-space) contribution directly using a direct Fourier transform, taking into account the offset of the subgrid in $uv$-space. After calculating the contribution of all visibilities within the subgrid, a fast Fourier transform (FFT) is used to transform each subgrid from image domain to $uv$-space, and the contribution of all the subgrids are added to the global $uv$-plane. Finally, the full $uv$-plane is transformed into the image using a FFT.

While this method performs more computations compared to convolutional gridding, it can be executed in parallel and is highly efficient when using graphics processing units (GPUs), resulting in a high gridding throughput. \textsc{idg} has been shown to speed up the gridding by an order of magnitude compared to traditional gridding algorithms \citep{veenboer-gpuidg-2017}.

When \textsc{idg} is used, anti-aliasing and $w$-term corrections are applied in image space, and are evaluated directly. This implies that \textsc{idg} is not affected by some of the errors made in traditional gridding algorithms, such as the discretization of $w$-values and the discretized gridding kernel. When using \textsc{idg} to predict visibilities, it has been shown that \textsc{idg} has a higher per-visibility accuracy compared to the $w$-stacking algorithm of \textsc{wsclean} \citep{vandertol-idg-2018}. Most of the calculations within \textsc{idg} are calculated with 32-bit single precision floating point values (IEEE 754-2008).

The \textsc{idg} implementation allows additional gridding terms, such as $w$-terms, primary beam terms ($a$-terms) and other direction-dependent effects. Unlike the $a$-projection algorithm \citep{bhatnagar-aproj-2008, tasse-2013-awimager, bhatnagar-2013-wb-aprojection}, the kernels are applied as multiplications in image space. Primary beam corrections could be important in the context of EoR experiments, in particular to correct for instrumental polarization leakage \citep{asad-2015, jagannathan-2017}. This is critical in power spectrum estimation \citep{jelic-2008} and for tomography with the SKA \citep{mellema-2015}. Full $a$-correction also allows per-station beam weighting during imaging. This allows an optimally weighted integration of the data. We will not focus on the errors associated with including such corrections, and instead limit the scope of this article to the gridding errors involved in the calculation of $w$-term corrected images without other direction-dependent effects.

\textsc{idg} is open source and available under the GNU General Public License v3.0.\footnote{The \textsc{idg} software is available at \url{https://gitlab.com/astron-idg/idg}.}, and has been integrated into \textsc{wsclean}. Therefore, \textsc{idg} can be combined with the deconvolution algorithms implemented in \textsc{wsclean}, such as the auto-masked multi-scale multi-frequency deconvolution algorithm \citep{offringa-2017}.

We use the \textsc{idg} default settings, which includes an optimized anti-aliasing kernel as described in \citet{vandertol-idg-2018}. For our setup, \textsc{idg} selects a subgrid size of 40 $\times$ 40 elements. IDG employs $w$-stacking to keep the size of the kernel, trimmed at the 1\% level, within the subgrid size. There is no oversampling parameter in IDG, because IDG always calculates the contribution of a visibility in real space, which implies there is no discretization of the $uv$-kernel.

\subsection{$w$-projection gridding} \label{sec:w-projection}
The $w$-projection algorithm applies the $w$-correction as a convolution in $uv$-space \citep{wprojection-cornwell}. For applying the $w$-projection algorithm, we use the \texttt{tclean} task in \textsc{casa} version 5.1.1-5 \citep{casa-2007}. The $w$-projection algorithm shares many of the configurable parameters of $w$-stacking, such as oversampling, anti-aliasing kernel size and padding, but these are not exposed in the \texttt{tclean} interface, and we will therefore use the default values: a prolate spheroidal kernel of $3 \times 3$, oversampling of a factor of 4 and a padding factor of 1.2. As with $w$-stacking, the $w$-direction needs to be discretized for $w$-projection, in order to precalculate a limited set of the $w$-kernels, and this leads to a similar parameter that sets the number of discretized $w$-values (the \texttt{wprojplanes} parameter in \textsc{casa}). In our analysis, we use \texttt{wprojplanes=256}. Furthermore, $w$-projection limits the $w$-kernel to a specific size, typically to the size at which the power goes below 1\% of the peak power \citep{wprojection-cornwell}.

\section{Power spectra} \label{sec:power-spectra}
21-cm power spectra quantify the spatial and spectral fluctuations found in the data. In this work, we calculate the power spectrum values from image cubes. The calculations follow those described in \citet{offringa-2019}, and consist of the following steps: i) spatial Fourier transformation; ii) normalization of the $uv$-values by dividing out the instrumental $uv$-response and converting them to kelvin; iii) a generalized inverse-variance weighted (with a diagonal matrix) least-squares Fourier transform along the line-of-sight direction; iv) cylindrical or spherical averaging. We will analyse data in two ways:
\begin{itemize}
 \item Using a foreground avoidance strategy. We will measure the power bias caused by gridding inside the foreground-free EoR window of a cylindrically-averaged power spectrum. In this approach, the modes inside the foreground wedge are not used.
 \item Using a foreground removal strategy. We use Gaussian process regression (GPR; \citealt{mertens-2018}) to remove residual foregrounds after gridding, and analyse the resulting full power spectra.
\end{itemize}
A Blackman-Harris window is used both during the spatial Fourier transform and during the least-squares inversion along the line of sight. We calculate the power spectra for baselines sizes of 50--250$\lambda$, corresponding with $k_\perp$-values of approximately 0.05--0.3~$h$Mpc$^{-1}$. These same settings are used in the analysis of LOFAR EoR observations \citep{patil-2017, mertens-10night-eorlimit-2019}.

\section{Simulated data} \label{sec:data}
To analyse the gridding accuracy we simulate a typical EoR observation with point sources drawn from a realistic population distribution. We use a distribution determined from low-frequency (154~MHz) observations \citep{franzen-2016}:
\begin{equation}
\frac{dN}{dS} = 6998\,S^{-1.54} \textrm{Jy}^{-1} \textrm{Sr}^{-1}.
\end{equation}
Using this distribution we predict sources with intrinsic (i.e., before applying the primary beam) flux densities between 1~mJy and 10~Jy in an area with a diameter of 90\degree, resulting in a model of approximately one million sources. We assume that all sources with an apparent flux density (i.e., after multiplying each source with the corresponding primary beam response) of at least 100 mJy can be subtracted from the visibilities before gridding, which is realistic for LOFAR observations: in LOFAR EoR observations the residual peak flux after direction-dependent subtraction of bright sources, imaged with a maximum baseline of 250$\lambda$, is about 70~mJy in the NCP field \citep{yatawatta-2013} and 150~mJy in the 3C\,196 field (Offringa et al. in prep). Therefore, we evaluate the average LOFAR primary beam value for each source and remove sources with an apparent brightness $>100$~mJy. To further limit the number of sources to be predicted, we also remove sources with an apparent flux density $<500$~$\mu$Jy, resulting in a model with $\sim$15,000 sources that are distributed out to 45{\degree} away from the phase centre. The spectral index of each source is drawn from a normal distribution with an average spectral index of $\alpha=-0.8$ (with $\alpha$ defined by $S(\nu)=S_0 (\nu/\nu_0)^\alpha$) and a standard deviation of 0.2. These distribution parameters match those of the weakest sources found by \citet{hurley-walker-2017-gleam}. We do not specifically simulate flattening of fainter (starburst) galaxies or special classes of sources such as USS, CSS or GPS sources that can have steep or curved spectra at the frequencies of interest (see \citealt{callingham-2017} for an overview).

The standard LOFAR software tool \textsc{dppp}\footnote{\textsc{dppp} is available at \url{https://github.com/lofar-astron/DP3}.} is used to predict fully accurate visibilities from the model by analytical evaluation of the visibility function and primary beam model. The observing time, phasing centre and antenna positions are taken from a 6~h night-time 3C\,196 observation. Besides gridding, several other processing steps can cause excess power, such as missing data due to RFI excision \citep{offringa-2019} and calibration with an incomplete model \citep{patil-2016, barry-2016, sardarabadi-2018}. In this work we limit ourselves to the effects of gridding, and therefore predict perfect data without flags or calibration errors. We do however include missing channels in our simulation, which are unavoidable in LOFAR data due to channel aliasing at the sub-band edges. The same effect also causes the sub-band edge channels of the Murchison Widefield Array (MWA; \citealt{tingay-2013-mwa}) to be lost \citep{offringa-2015-mwa-rfi}. In LOFAR EoR processing, two 3~kHz channels at each side of the sub-band are removed before averaging, leaving 60/64 channels in the data for each 195~kHz sub-band. These data are averaged by a factor of 12 in frequency and 6 in time, resulting in 12~s timesteps and 5 channels per sub-band, with gaps between the sub-bands. The decorrelation caused by averaging is $\ll 1\%$. In this work, we directly forward predict the averaged data, and are therefore not affected by time or frequency smearing. We simulate data between 115--134~MHz, 94 sub-bands in total, each with 5 channels.

\section{Results} \label{sec:results}
To assess the effects of gridding, we independently image each of the 470 frequency channels of our simulated data (\S\ref{sec:data}) using \textsc{wsclean}, and construct 21-cm power spectra from the resulting image cube. These power spectra are compared to ground-truth power spectra that are made from the direct-FTed images.

The images cover $3${\degree} by $3${\degree} on the sky with $360 \times 360$ pixels. Our limited imaging field of view implies that only the most sensitive part of the primary beam is used. In the corners of the images, the beam has a gain of approximately 75\%.

\begin{figure*}[p!]
\centering
\includegraphics[width=90mm]{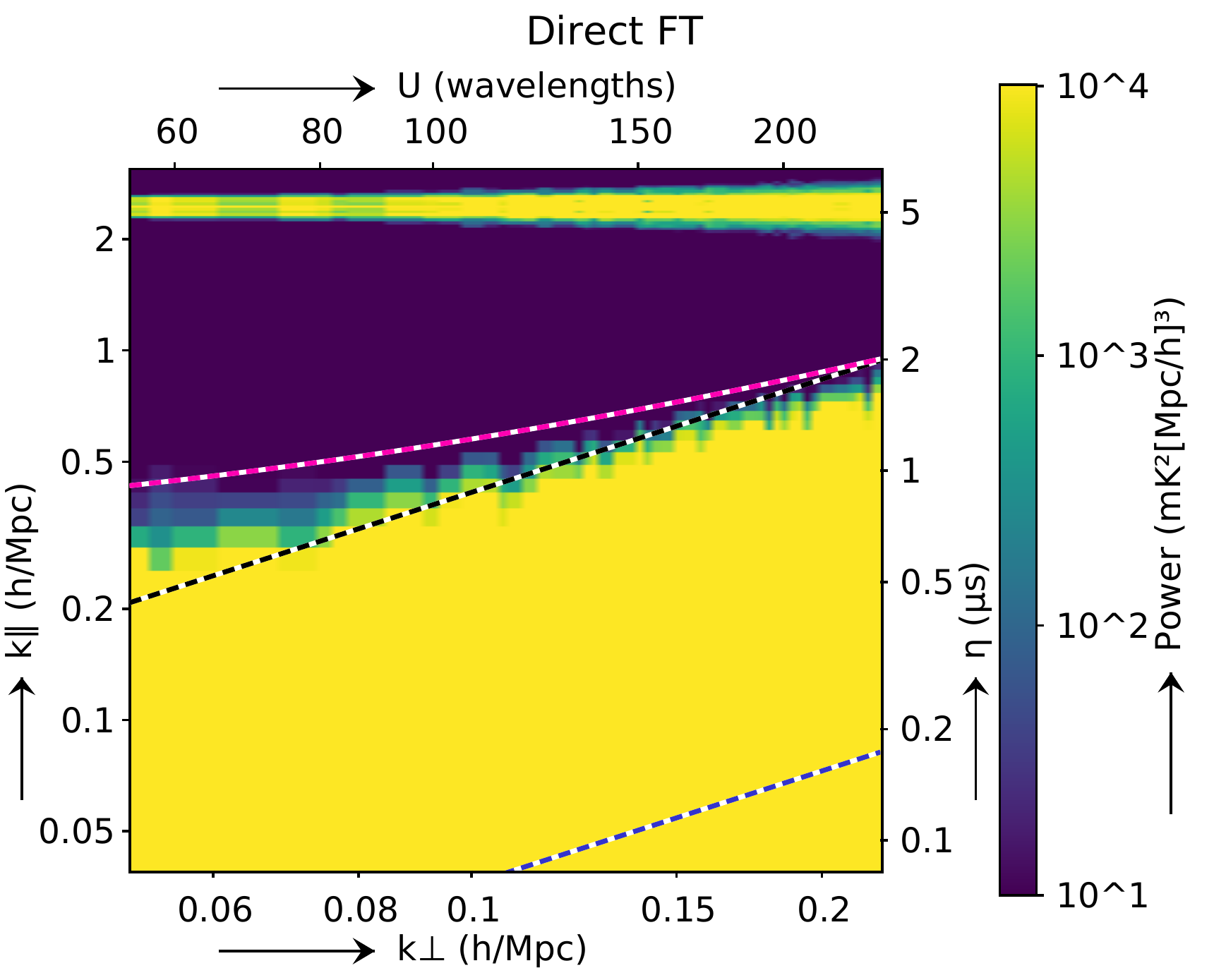}%
\includegraphics[width=90mm]{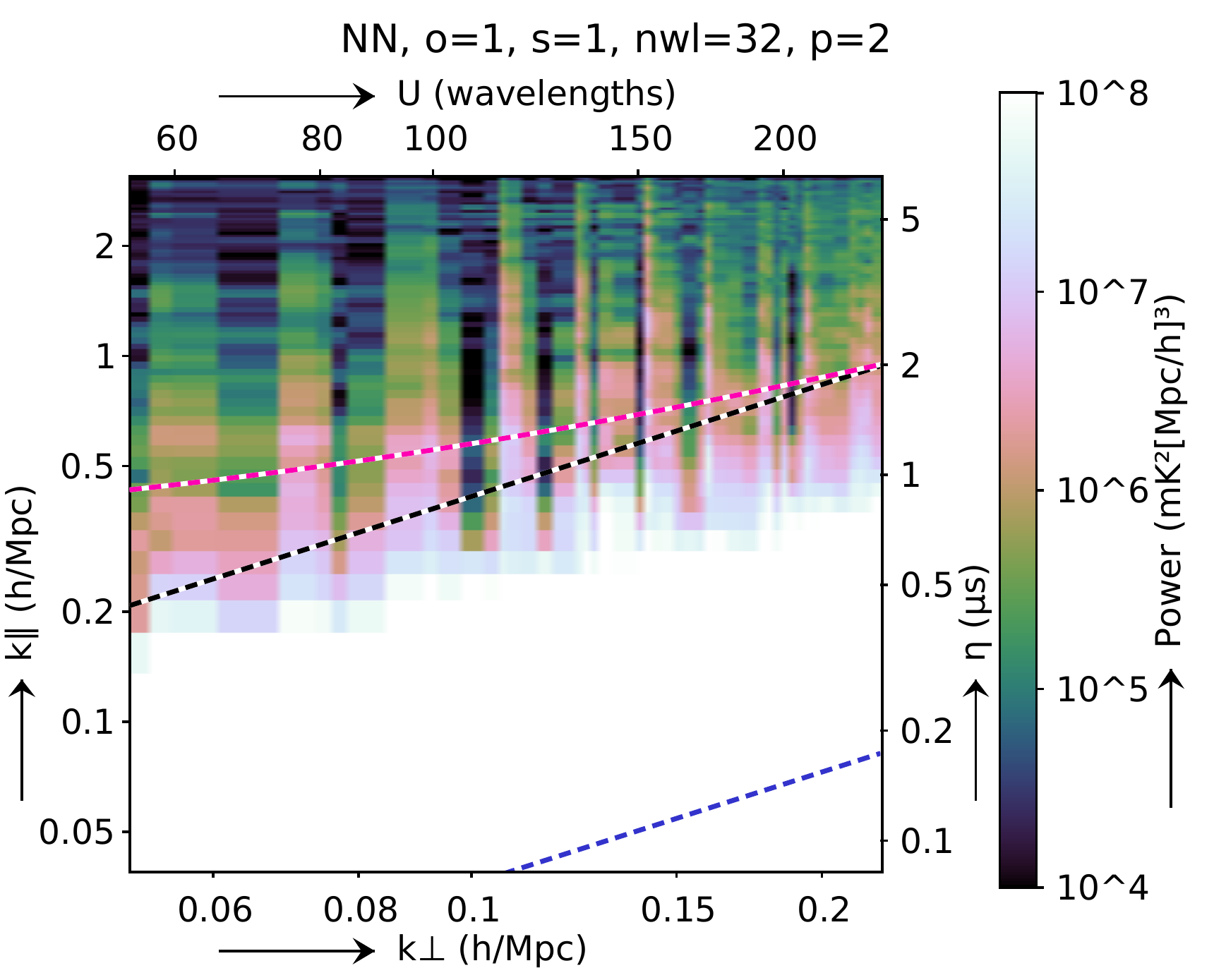}\\%
\includegraphics[width=90mm]{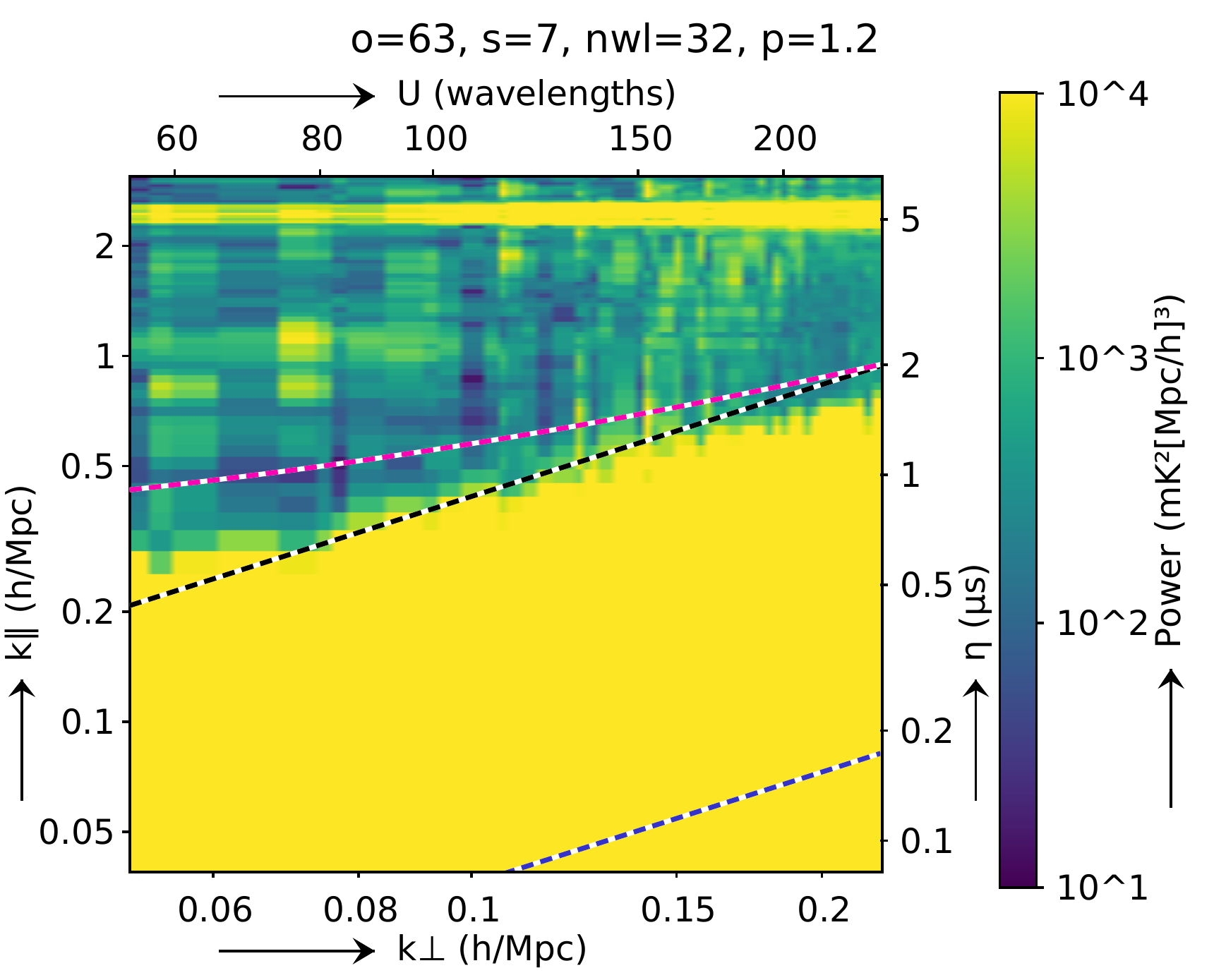}%
\includegraphics[width=90mm]{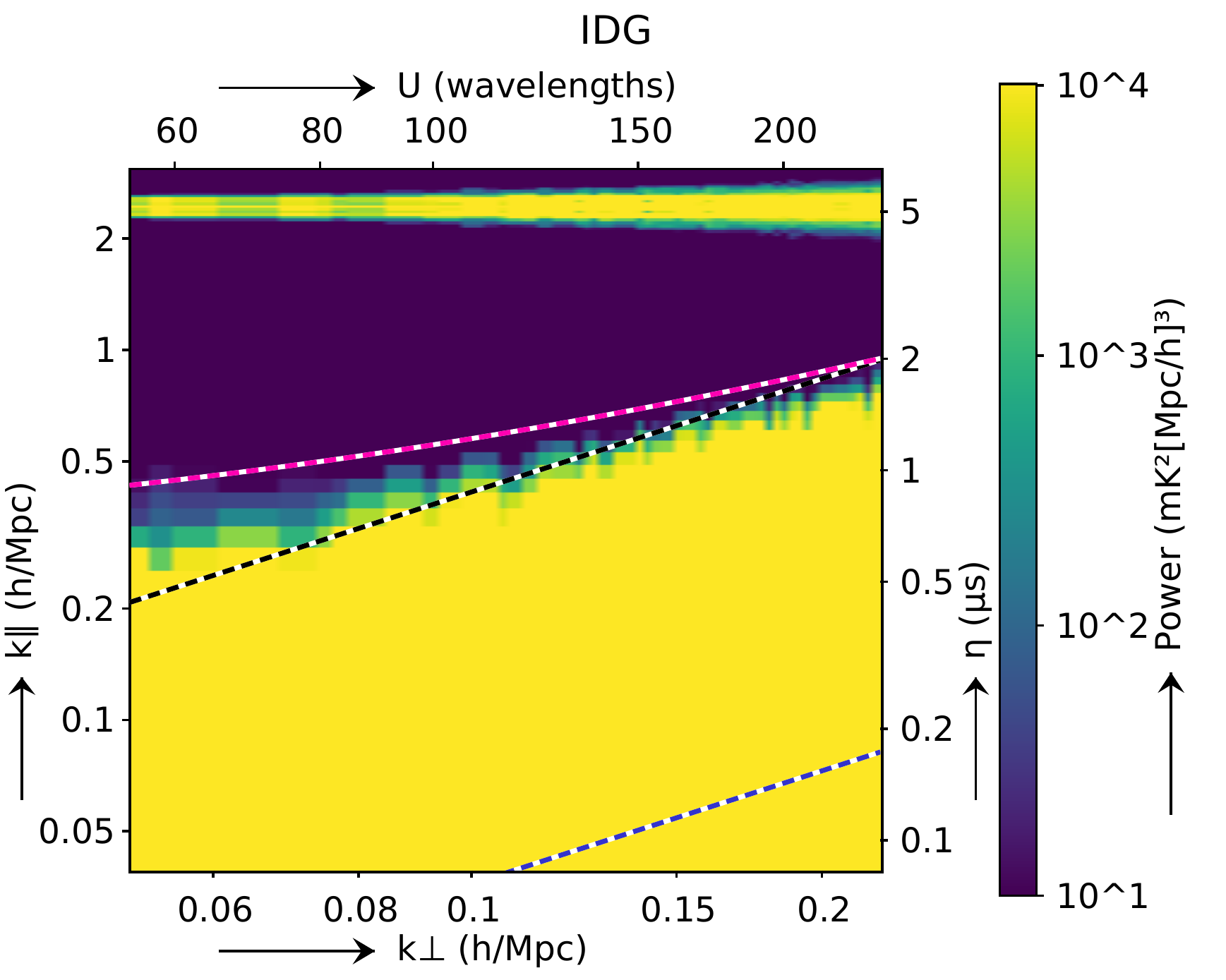}\\%
\includegraphics[width=90mm]{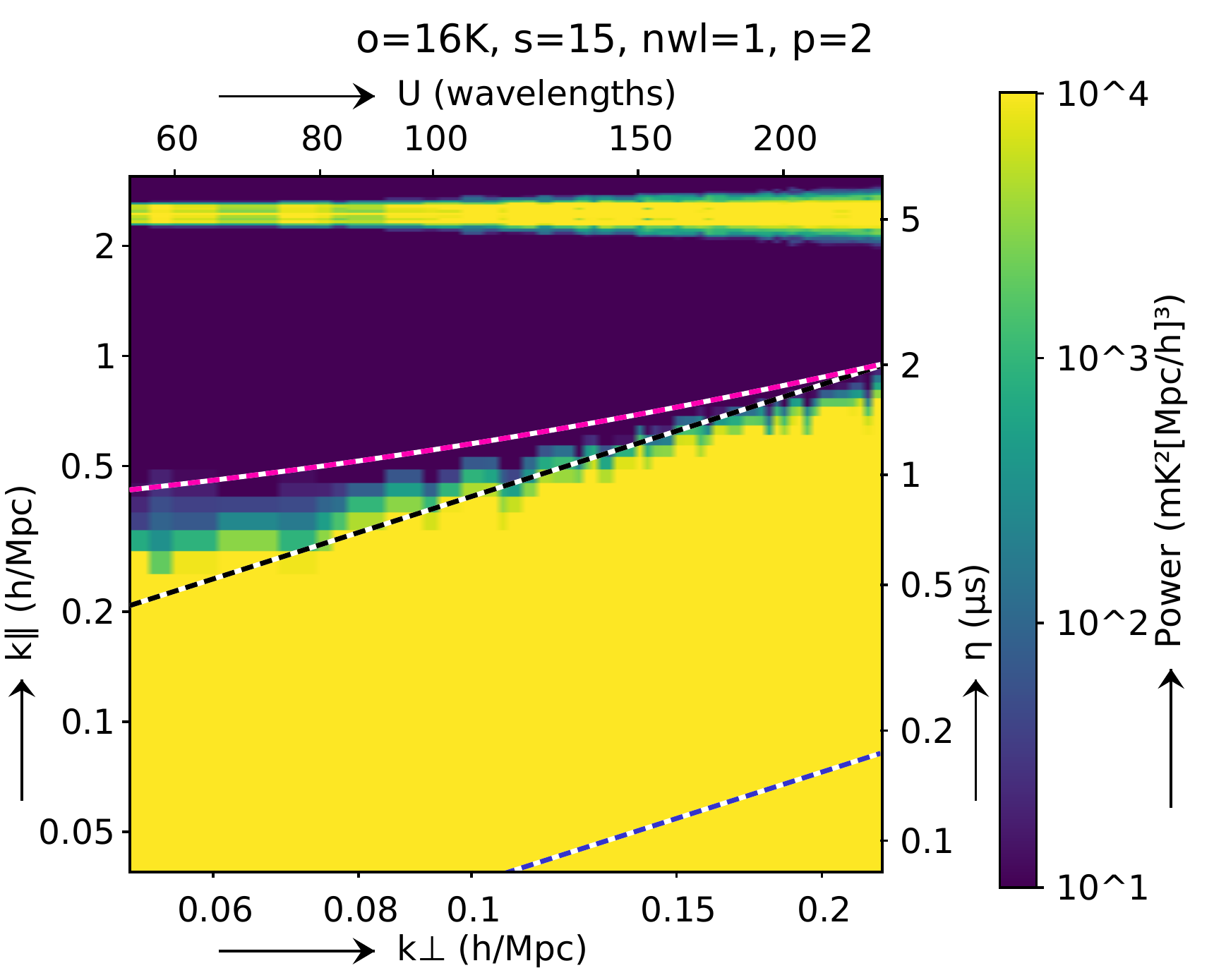}%
\includegraphics[width=90mm]{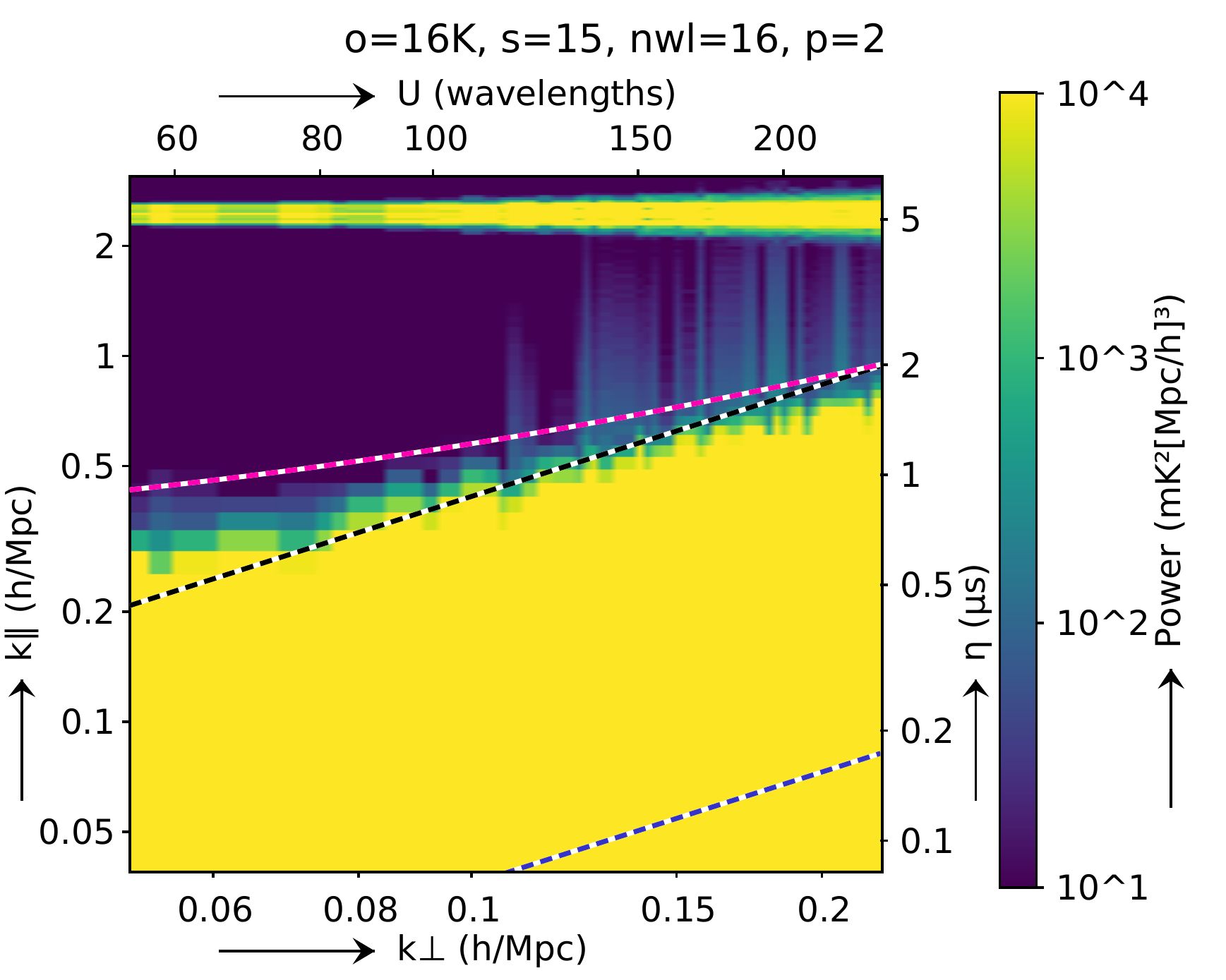}%
\caption{Cylindrically-averaged power spectra for various gridding settings. From left to right, top to bottom: direct FT inversion; nearest neighbour gridding (no oversampling); default settings for \textsc{wsclean}; default settings for image domain gridding; increased oversampling and kernel size settings for \textsc{wsclean}, without $w$-correction; and including $w$-correction but with a low number of $w$-layers. Nearest-neighbour gridding results are drawn with a different colour scale. Black dashed line: horizon wedge; pink dashed line: same with extra space for windowing function; blue dashed line: the primary beam (5\degree) wedge. Gridding parameters are abbreviated as follows: o = oversampling factor; s = gridding kernel size; nwl = number of $w$-layers; p = padding factor.}
\label{fig:power-spectra-2d-before-gpr}%
\end{figure*}

\begin{figure*}[htb]
\centering
\includegraphics[width=15cm]{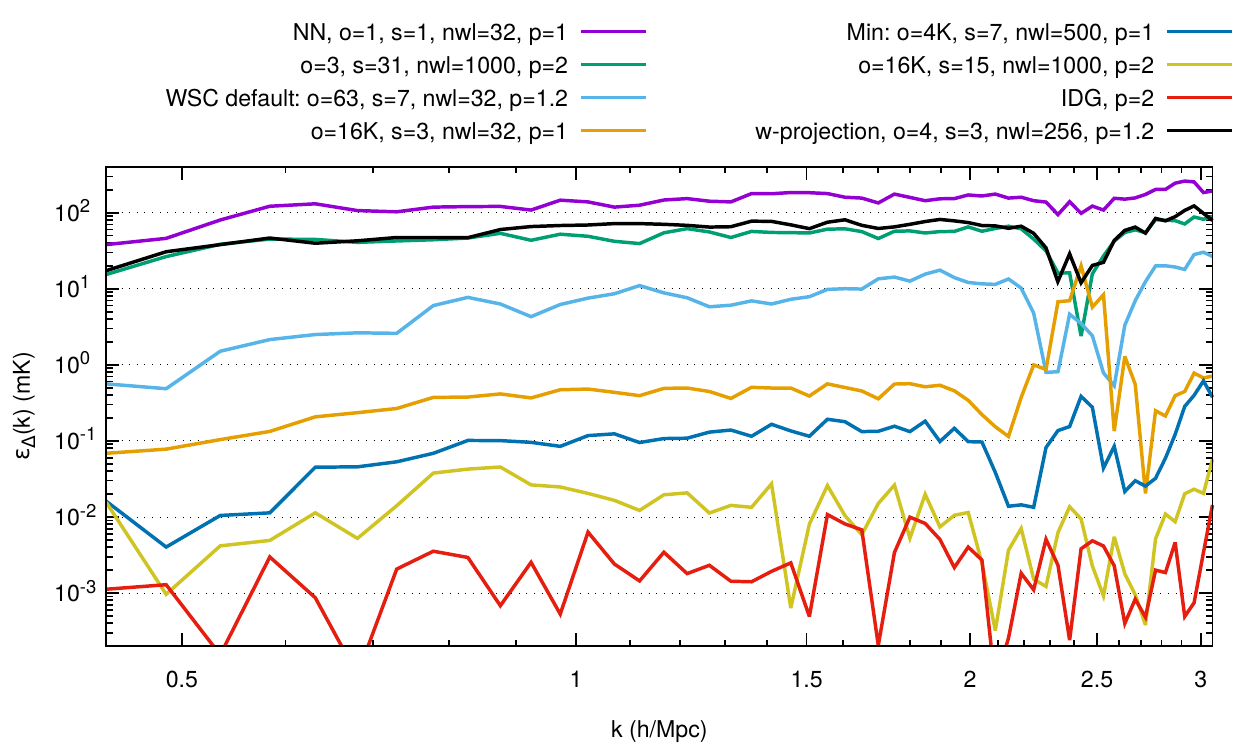}%
\caption{Spherically-averaged ``foreground avoidance'' power spectra errors (absolute difference) without GPR foreground subtraction, compared to the directly FTed data. $k$-values that fall under the wedge are excluded. Gridding parameters are abbreviated as follows: o = oversampling factor; s = gridding kernel size; nwl = number of $w$-layers; p = padding factor.}
\label{fig:powerspectrum-1d-before-gpr}%
\end{figure*}

\subsection{Foreground avoidance results}
We start by investigating a foreground avoidance strategy. In this scenario, the modes that are dominated by foregrounds are not used in the final power spectra, and we therefore do not perform Gaussian progress regression to remove the wedge. Before performing the $k_\parallel$ transform, a third-order polynomial fit in frequency direction is subtracted from the $uv$-cube, from the real and imaginary parts separately. This removes both EoR and foreground power from the low $k_\parallel$-modes inside the wedge. This decreases the dynamic range requirements of the generalized $k_\parallel$ Fourier transform, thereby avoiding some artefacts that are not the focus of this paper, without biasing the power spectrum in the parts that we measure.

Fig.~\ref{fig:power-spectra-2d-before-gpr} shows cylindrically-averaged power spectra for various gridding methods to provide an overview of the artefacts that each method produces. The foreground wedge structure is clearly visible. Power under the wedge is saturated in the colour scale used in these plots. The strongest modes within the wedge have values of $10^{11}$~mK$^2h^{-3}$Mpc$^3$, which implies a dynamic range of over ten orders of magnitude between contaminated and uncontaminated modes. A horizontal line at $k=2.4$~$h$Mpc$^{-1}$ (delay of 5~$\mu$s) is caused by the spectral gap between sub-bands. Fig.~\ref{fig:power-spectra-2d-before-gpr} demonstrates that gridding can cause different artefacts in the 2D power spectra: excess power that is strongest at low $k_\parallel$-values (nearest-neighbour gridding), a uniform level of excess power (default \textsc{wsclean} settings: $w$-stacking with 32 $w$-layers, kernel size of 7, 1.2$\times$ padding, 63$\times$ oversampling), and excess power at the longest baselines (limited $w$-sampling).

An overview of the effect of various settings in a spherically-averaged power spectrum is given in Fig.~\ref{fig:powerspectrum-1d-before-gpr}. Only modes outside the wedge are integrated. We add a delay of 0.6~$\mu$s to the theoretical horizon wedge line to also exclude the convolution kernel size resulting from the windowing in the $k_\parallel$-transform, resulting in the pink dashed line in Fig.~\ref{fig:power-spectra-2d-before-gpr}. When comparing the different methods by their excess power above the wedge, nearest-neighbour gridding results in strong excess power, with an excess of about 100~mK. The $w$-projection implementation in \textsc{casa} shows an excess of $\sim$20-50~mK. Both exceed nearly all 21-cm EoR models. With the default \textsc{wsclean} settings, this decreases to 1~mK at low $k$-values and to 10~mK at high $k$-values. Gridding with \textsc{idg} results in very accurate results with the least excess power (1 to 10~$\mu$K) of all tests.

\begin{figure*}
\centering \vspace*{-5mm} %
\hspace*{-5mm}\includegraphics[width=115mm]{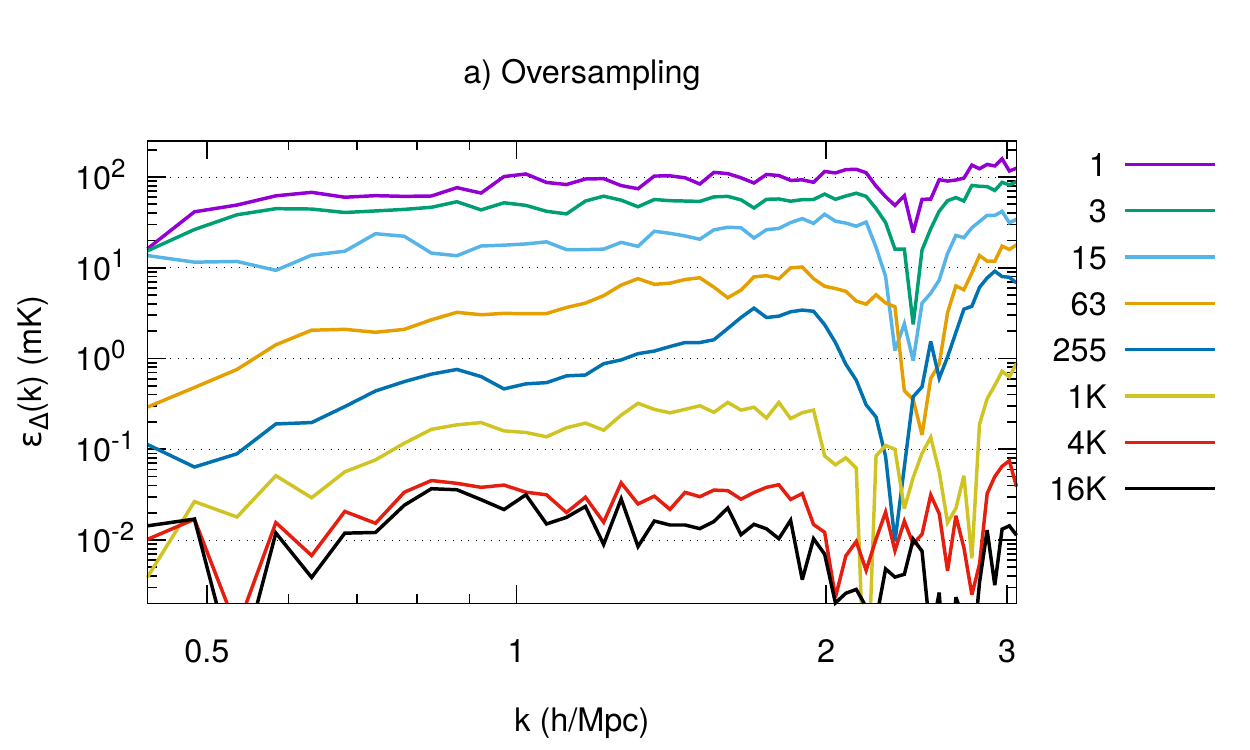}\\%
\hspace*{-5mm}\includegraphics[width=95mm]{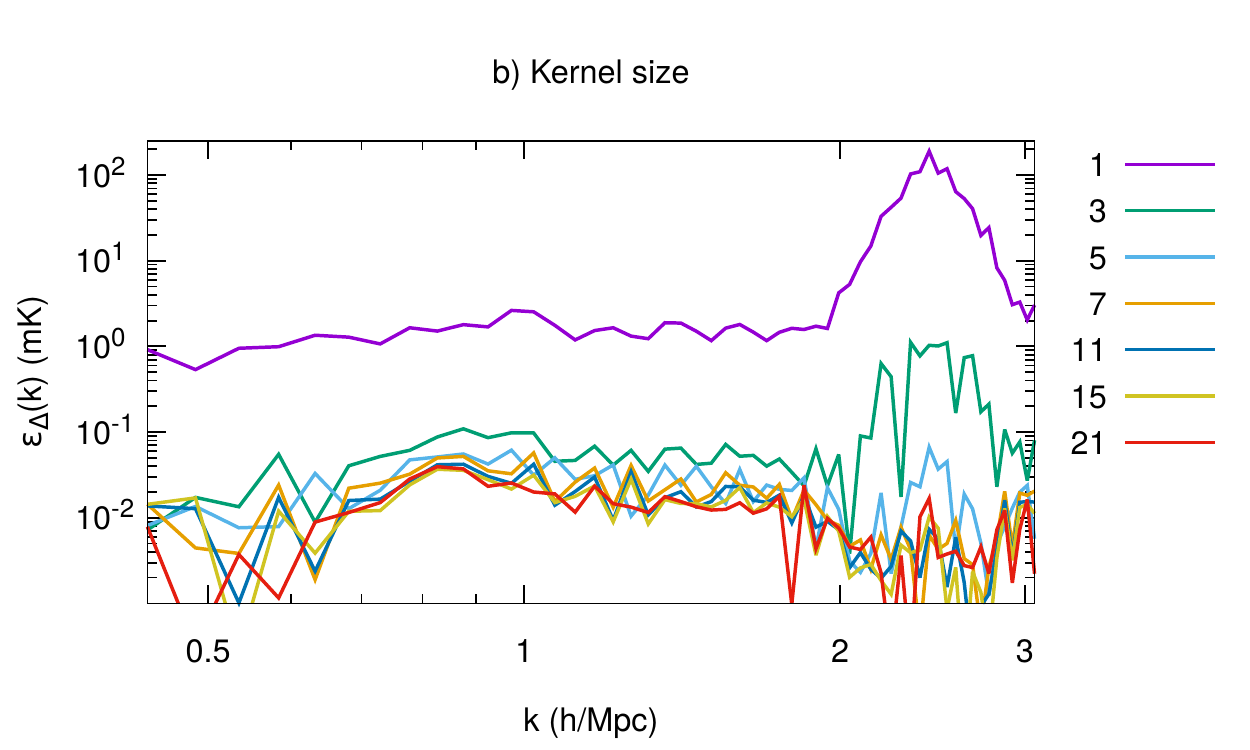}%
\includegraphics[width=95mm]{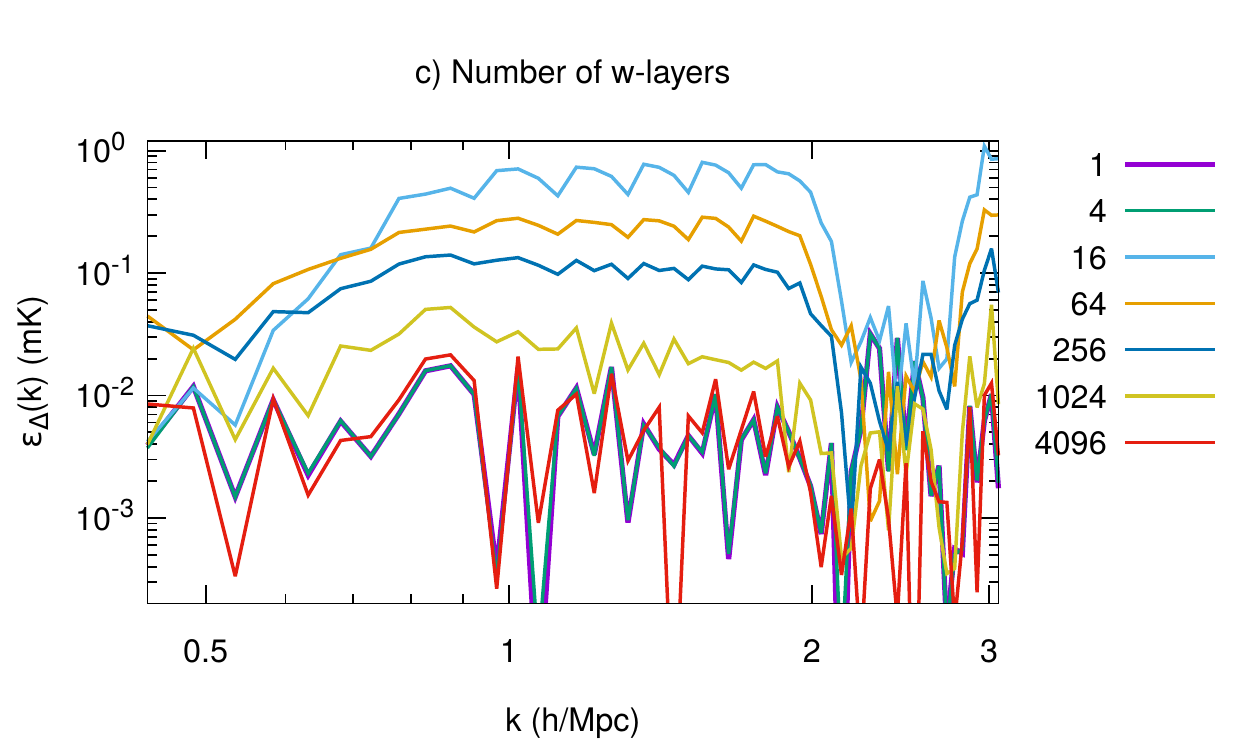}\\%
\hspace*{-5mm}\includegraphics[width=95mm]{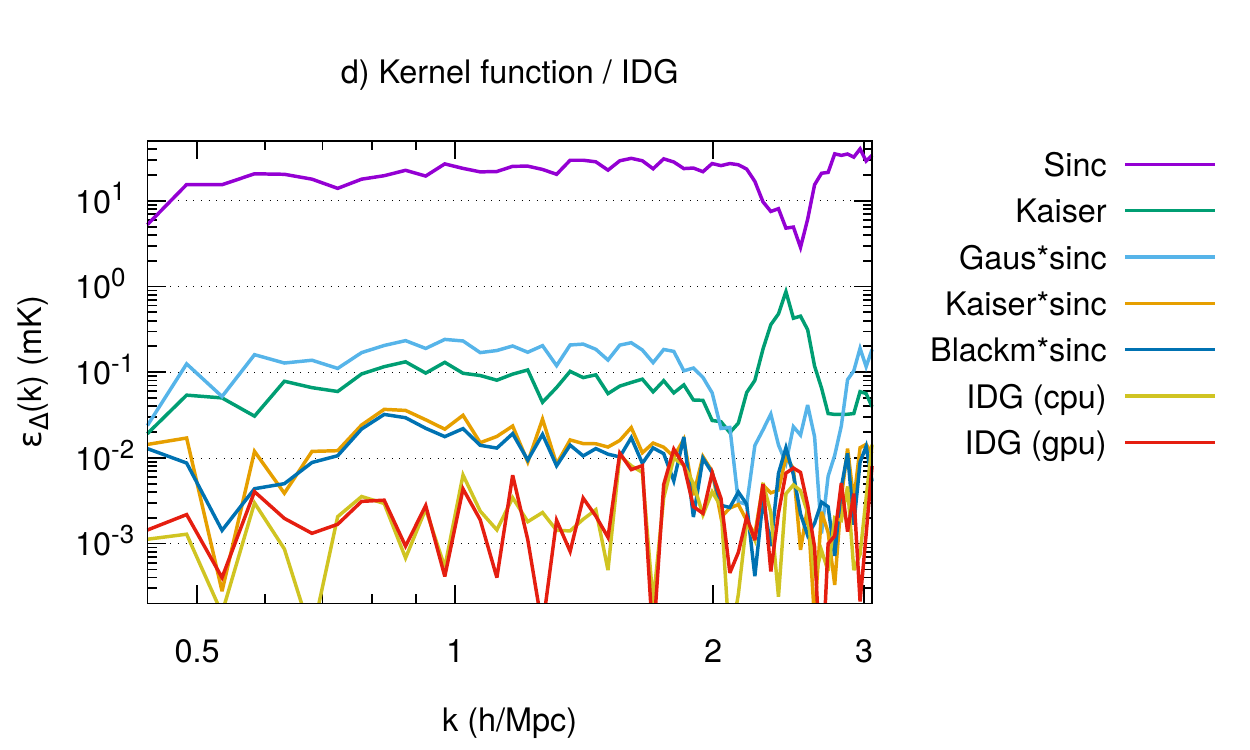}%
\includegraphics[width=95mm]{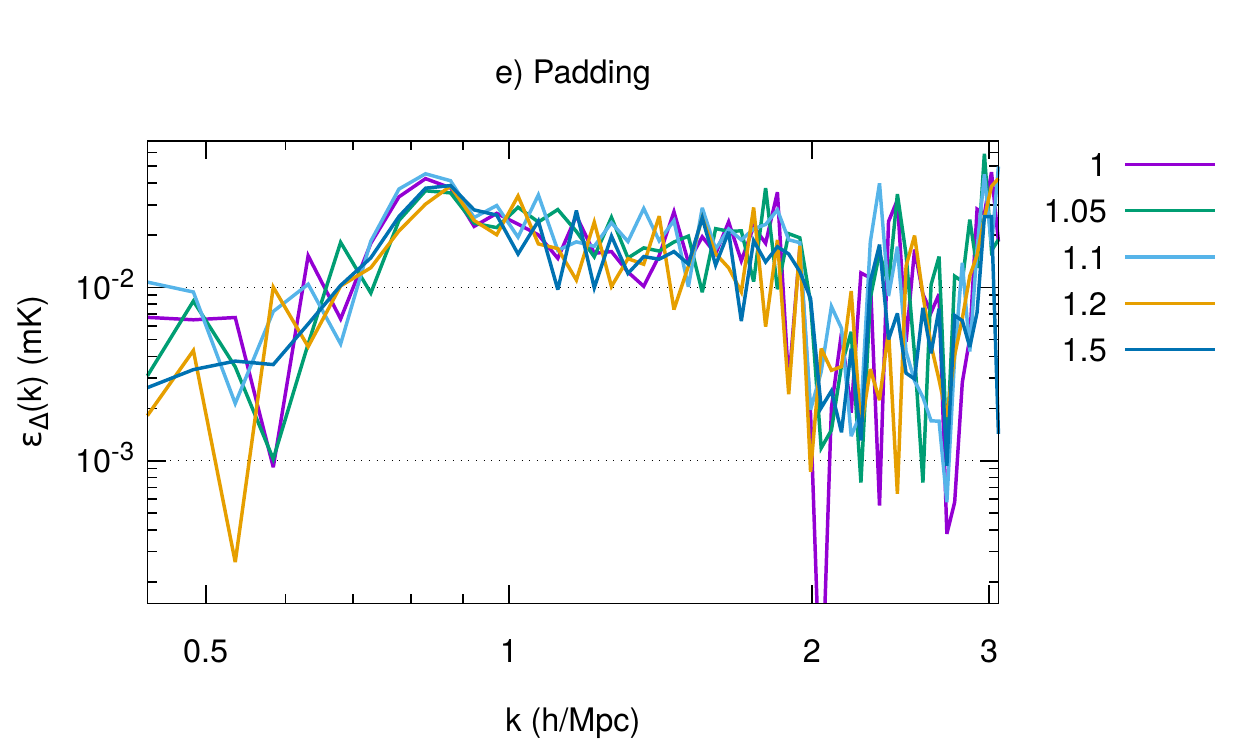}%
\caption{Effect of gridding accuracy for several gridding parameters: a) kernel oversampling (\S\ref{sec:results-oversampling}); b) kernel size (\S\ref{sec:results-kernelsize}); c) the number of $w$-discretization levels (\S\ref{sec:results-wlayers}); d) gridding kernel function and \textsc{idg} comparison (\S\ref{sec:results-kernelfunction} and \S\ref{sec:results-idg}); and e) padding (\S\ref{sec:results-padding}). The plots describe the absolute error of spherically-averaged power spectrum measurements using a foreground avoidance strategy. The direct FT results are used as ground truth. Each plot shows the dependency on one parameter, while keeping the other parameters at their highest accuracy setting (see Table~\ref{tbl:parameters}). }
\label{fig:power-spectra-per-parameter}%
\end{figure*}

\begin{table}
\caption{Gridding parameter values. Columns 2, 3 and 4 specify the default settings in \textsc{wsclean}; the settings used in Fig.~\ref{fig:power-spectra-per-parameter} (unless otherwise specified); and the minimum settings that are required to have an excess power of at most 0.1~mK in the range $k$=0.5--1~$h$Mpc$^{-1}$, respectively. The latter holds for both the foreground avoidance and the foreground subtraction approach.}
\label{tbl:parameters}
\begin{tabular}{|l|c|c|c|}
\hline
\textbf{Name} & \textbf{Default} & \textbf{Fig.~\ref{fig:power-spectra-per-parameter}} & \textbf{Minimum} \Tstrut\Bstrut \\
\hline
Oversampling & 63 & 16535 & 4095 \Tstrut \\
Kernel size & 7 & 15 & 3\\
$w$-layers & 32 & 1000 & 500 \\
Kernel function & Sinc$\times$KB & Sinc$\times$KB & Sinc$\times$KB \\
Padding & 1.2 & 2 & 1 \Bstrut \\
\hline
\end{tabular}
\end{table}

Fig.~\ref{fig:power-spectra-per-parameter} shows various foreground-avoiding power spectra, each visualizing the result of changing the value of one parameter while the other parameters are fixed to a setting that reflects a high accuracy for that parameter. For each parameter, we will determine the least computationally expensive setting that would still allow a detection of the 21-cm signals from the Epoch of Reionization. The 21-cm signals are expected to have a brightness of a few mK (e.g., \citealt{greig-2015-21cmmc, mellema-2018}), and we therefore require that less than 0.1~mK power is added in the range of $k=0.5$--$1$~$h$Mpc$^{-1}$. The parameter settings are summarized in Table~\ref{tbl:parameters}.

\subsubsection{Oversampling} \label{sec:results-oversampling}
The results indicate that the oversampling factor is the most crucial parameter for avoiding gridding excess power. 
Fig.~\ref{fig:power-spectra-per-parameter}a shows that the default setting of 63 for \textsc{wsclean} adds a few mK excess power. Therefore, the default settings do not meet the minimum accuracy. Oversampling with a factor of $4\times10^3$ limits the excess noise below 0.1~mK
(34~$\mu$K at $k=1$~$h$Mpc$^{-1}$).
With an oversampling of approximately $8\times 10^3$ times, the excess power is no longer reduced by increasing the oversampling further, indicating that the error due to sampling of the kernel is no longer the limiting factor. The added computational cost of increasing the oversampling factor is relatively small because the gridding kernel is precalculated. Increasing the oversampling from 63 to $8\times 10^3$ increases the imaging time by less than 10\%. The need for large oversampling factors also explains why the $w$-projection result in Fig.~\ref{fig:powerspectrum-1d-before-gpr}, for which an oversampling factor of 4 is used, shows a high level of excess power.

\subsubsection{Kernel size} \label{sec:results-kernelsize}
As shown in Fig.~\ref{fig:power-spectra-per-parameter}b, a kernel size of 3 is enough to limit the excess noise below 0.1 mK at $k=1$ $h$Mpc$^{-1}$. This implies that the default size of 7 can be decreased for EoR experiments. However, decreasing the kernel size from 7 to 3 does not improve gridding speed \citep{offringa-wsclean-2014}.

\subsubsection{\textit{w}-layers} \label{sec:results-wlayers}
The bottom left figure of Fig.~\ref{fig:power-spectra-2d-before-gpr} shows the result of applying no $w$-term correction. This demonstrates that $w$-correction is not strictly required to avoid excess noise. However, the lack of $w$-correction causes some decorrelation to occur, which in turn reduces sensitivity. The amount of decorrelation is dependent on the image size, baseline length and array configuration. When disabling $w$-term correction, we measure a root mean square error of 9\% over the full image, and an average loss of 8\% in source strength at $1.5\degree$ distance for our imaging configuration (3$\degree$ $\times$ 3$\degree$ FOV, LOFAR baselines up to 250$\lambda$). Fig.~\ref{fig:power-spectra-per-parameter}c shows that using a small number of $w$-layers of for example 16 causes more excess power compared to using no $w$-layers at all. This can be explained by how $w$-stacking works: it groups visibilities with similar $w$-terms and uses a constant $w$-correction for those. Because the $w$-term is frequency dependent, whereas the maximum $w$-term (and therefore the $\Delta w$ stepsize) is limited by the baseline length threshold (250$\lambda$), this causes fluctuations over frequency. To avoid significant decorrelation and excess noise, at least 300 $w$-layers are necessary. Using 300 $w$-layers increases the imaging time by a factor of 3 compared to no $w$-correction.

\subsubsection{Kernel function} \label{sec:results-kernelfunction}
Fig.~\ref{fig:power-spectra-per-parameter}d shows the results for gridding with different kernel functions: a truncated sinc-function, the Kaiser-Bessel function, and a sinc windowed by a truncated Gaussian, Kaiser-Bessel and Blackman-Nutall function. Windows with stronger sidelobe suppression cause less excess power. This underlines that kernels with discontinuities at the border will cause spectral fluctuations.

\subsubsection{\textsc{idg}} \label{sec:results-idg}
In addition to different kernel functions, Fig.~\ref{fig:power-spectra-per-parameter}d also shows the \textsc{idg} results with CPU and GPU, which both show a low excess power of a few $\mu$K over most of the measured $k$-range. The two results show slightly different results, which might be caused by different implementations of the $\sin$ and $\cos$ functions or the use of a different fast Fourier transform library.

\subsubsection{Padding} \label{sec:results-padding}
Padding mitigates edge effects in the image domain. As demonstrated by Fig.~\ref{fig:power-spectra-per-parameter}e, padding has no significant effect on the gridding excess power in 21-cm analysis. This can be explained by the fact that the edge effects do not cause spectral fluctuations.

\subsubsection{Numerical precision} \label{sec:results-numprec}
We have compared a direct Fourier transform performed with single precision floats and with double-precision floats, and observe no significant differences between the two results. This suggests that gridding with single-precision floating point calculations is accurate enough for EoR experiments. In general, adding a large number of values together can result in a loss of precision, and with $2.6\times10^9$ visibilities that might seem inevitable. A reason why in practice we see no difference between single and double-precision floats, is that values in image space grow with the square root of the number of samples. In $uv$-space, values are naturally dispersed because they are gridded in different $uv$-bins.

\begin{figure*}
\centering
\includegraphics[width=8cm]{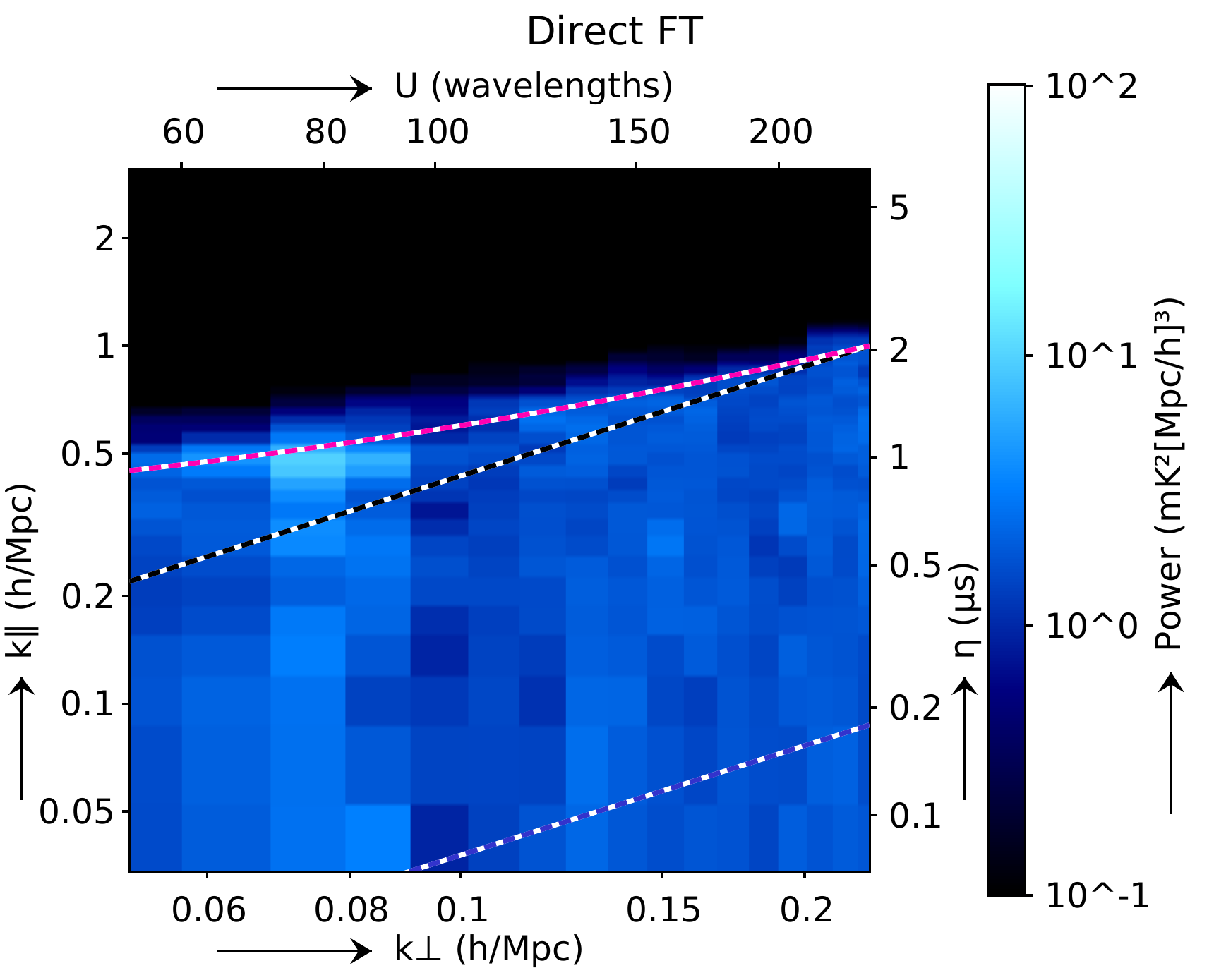}%
\includegraphics[width=8cm]{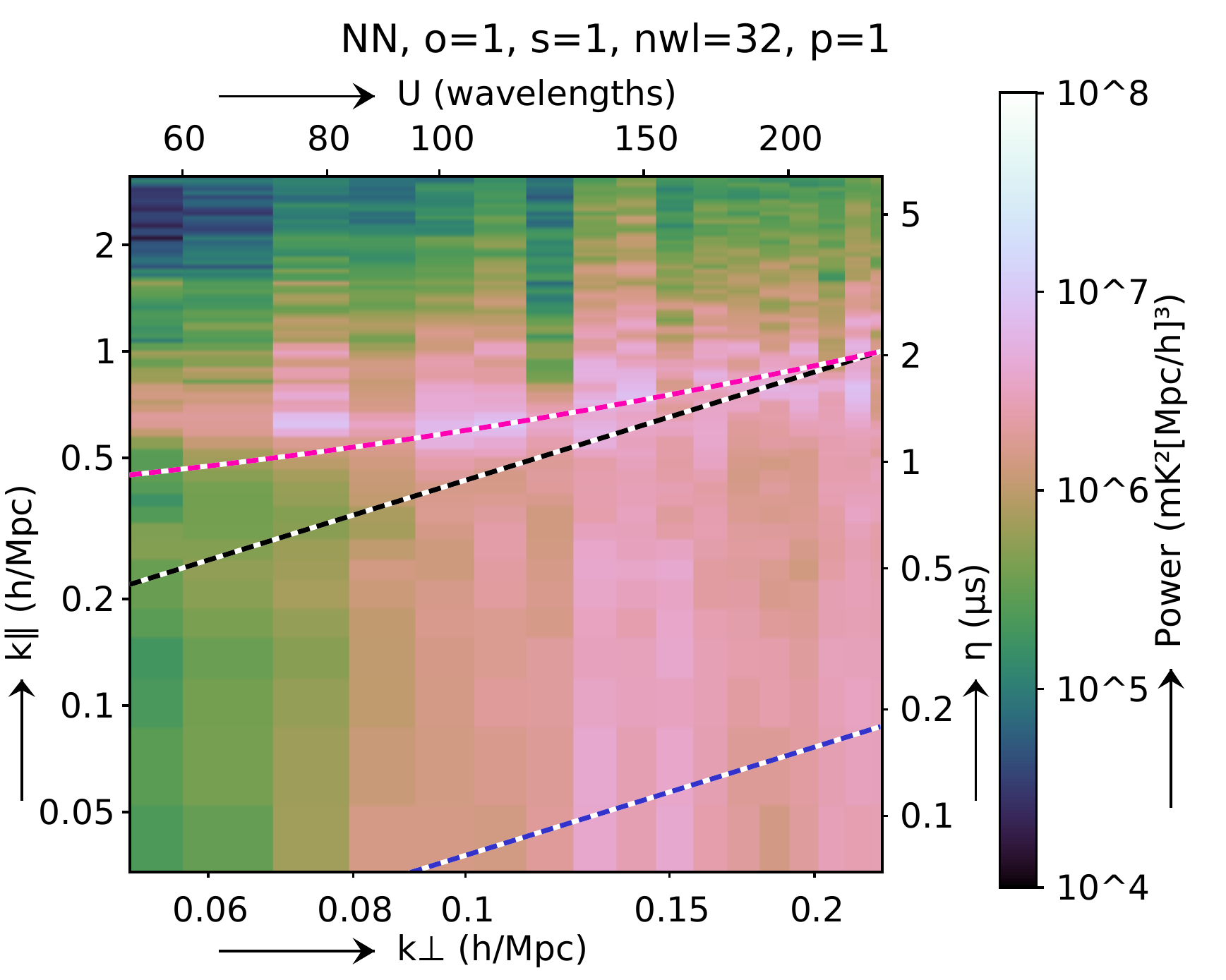}\\%
\includegraphics[width=8cm]{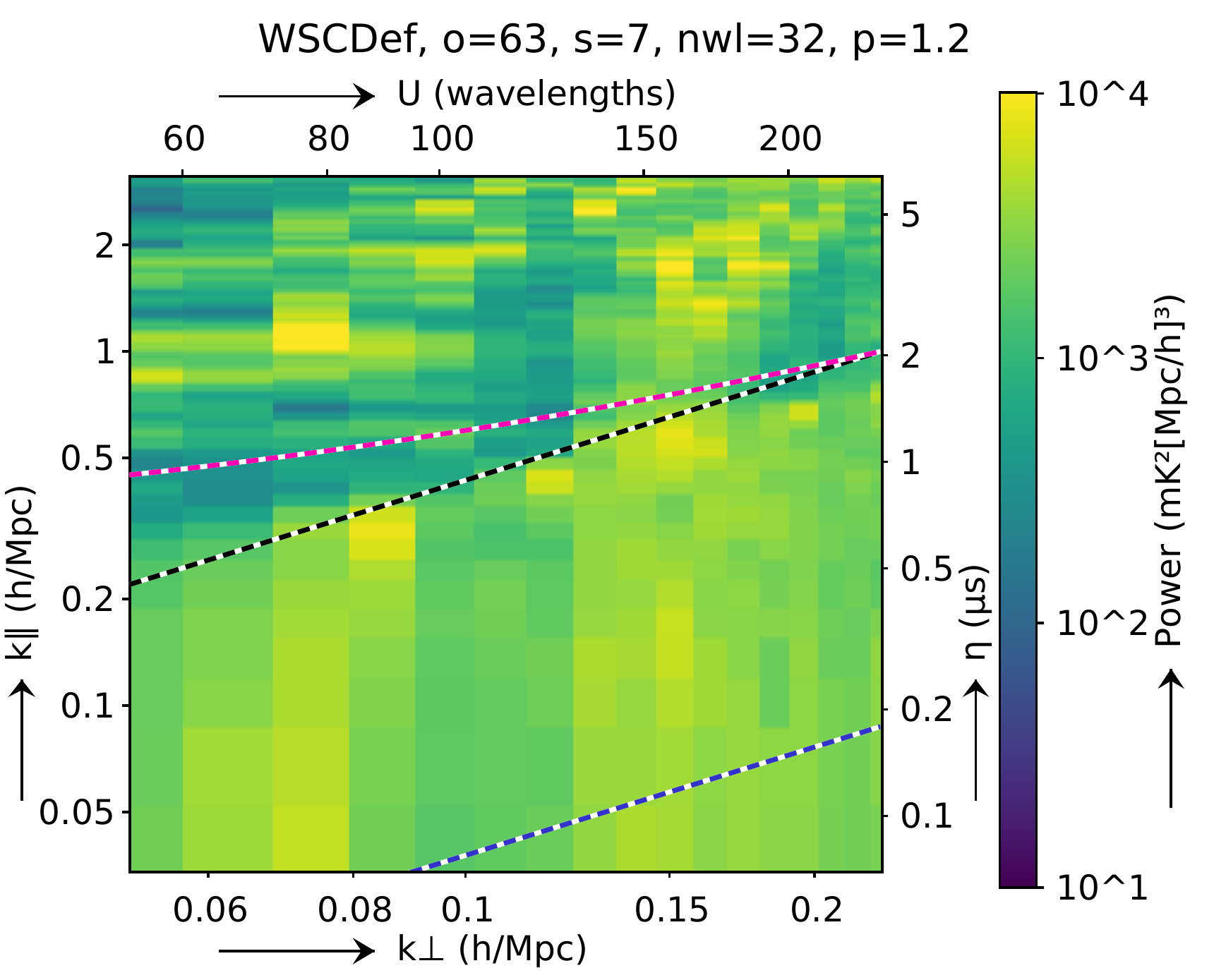}%
\includegraphics[width=8cm]{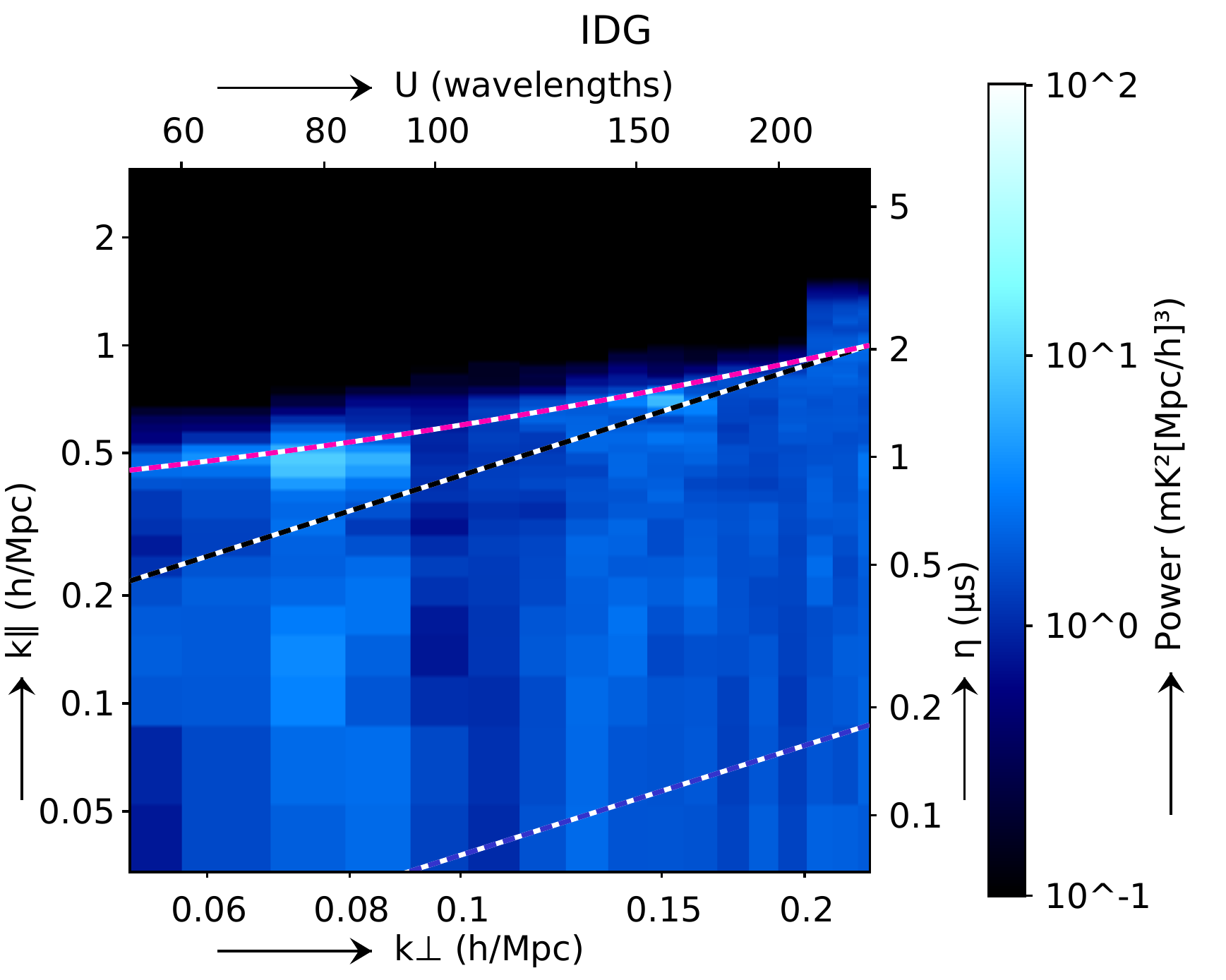}\\%
\includegraphics[width=8cm]{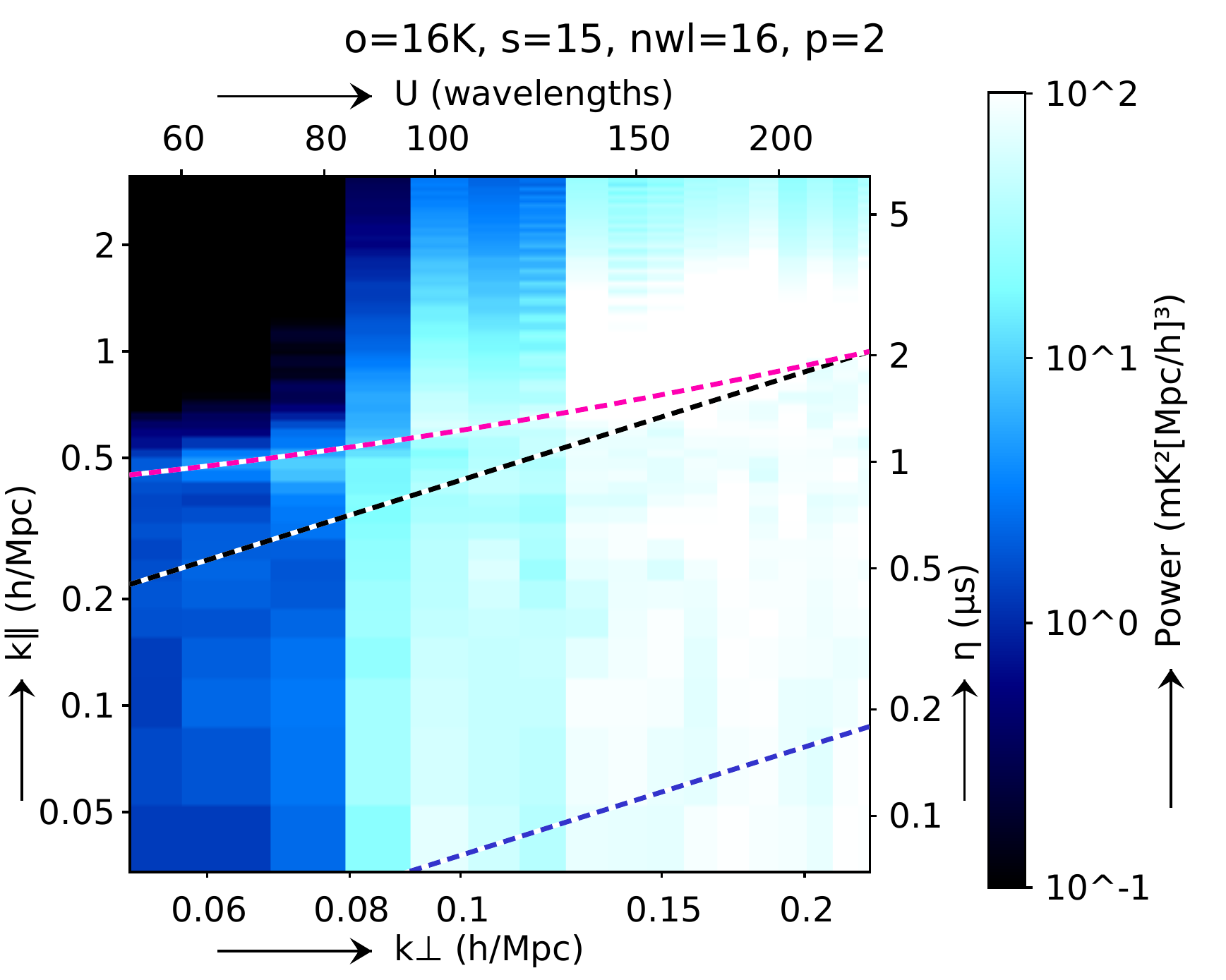}%
\includegraphics[width=8cm]{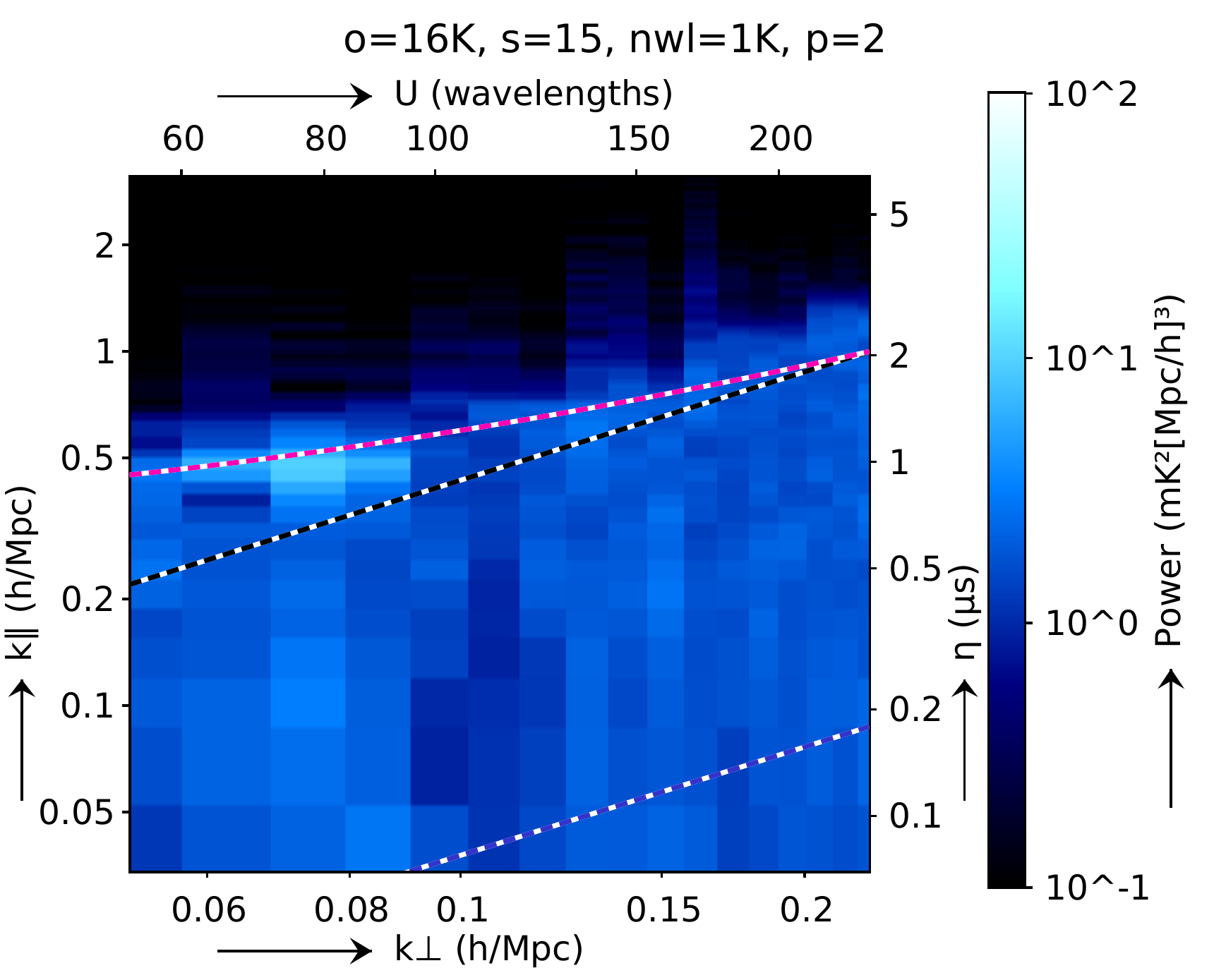}%
\caption{Residual cylindrically-averaged power spectra after applying GPR. From left to right, top to bottom: direct FT inversion; nearest neighbour gridding (no oversampling); default settings for \textsc{wsclean}; default settings for image domain gridding; increased oversampling and kernel size settings for \textsc{wsclean}; and same but with a high number of $w$-layers. To highlight the excess power, not all power spectra use the same color maps. Black dashed line: horizon wedge; pink dashed line: same with extra space for windowing function; blue dashed line: the primary beam (5\degree) wedge. Gridding parameters are abbreviated as follows: o = oversampling factor; s = gridding kernel size; nwl = number of $w$-layers; p = padding factor. }
\label{fig:power-spectra-2d-after-gpr}%
\end{figure*}

\begin{figure*}
\centering
\includegraphics[width=15cm]{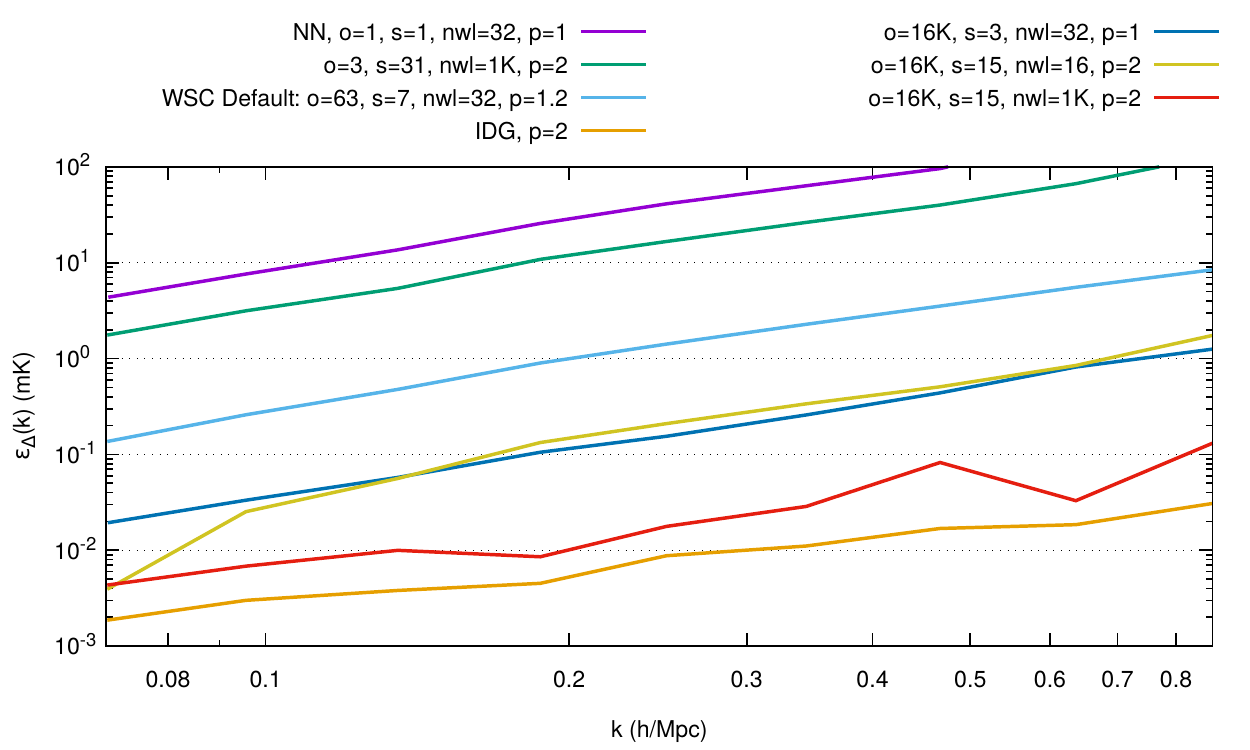}%
\caption{Spherically-averaged ``foreground subtraction'' power spectra errors (absolute difference) after foregrounds removal with GPR. The ground truth (power spectrum from directly FTed data) was subtracted from each resulting power spectra. All $k$-values are included. Gridding parameters are abbreviated as follows: o = oversampling factor; s = gridding kernel size; nwl = number of $w$-layers; p = padding factor.}
\label{fig:powerspectrum-1d-after-gpr}%
\end{figure*}

\subsection{Foreground subtraction results}
In this section, we discuss the results of applying Gaussian progress regression (GPR) to the data to remove the emission in the wedge, and subsequently including the foreground-contaminated modes in the power spectra.

GPR has the potential to cause some bias of the signal \citep{mertens-2018}. A full quantization of this bias is beyond the scope of this paper, but we made a simple simulation to test the performance of GPR with the settings and foregrounds that are used in this paper. This simulation consist of the predicted foregrounds with the most accurate gridding settings, a noise equivalent to 100 nights of 12~h and a realistic system equivalent flux density for LOFAR of 4000~Jy per station, and a 21-cm signal covering a large range of variances and frequency coherence scales.

For each of these signal strengths and coherence-scales, 10 realizations of noise and signal are generated and GPR is performed on the summed images. The ratio of input over recovered signal power-spectra is computed for three different ranges of scales. We find biases in the range 0.7--2.5, and overall similar results to what was found in \citet{mertens-2018}.

We continue by applying GPR to the foreground-only image cubes with different gridding settings, and construct power spectra from the GPR residuals. Fig.~\ref{fig:power-spectra-2d-after-gpr} shows the cylindrically-averaged power spectra after having removed the foregrounds with GPR. In the direct FT result, the residual foreground power is about 2 mK$^2h^{-3}$Mpc$^3$, a factor of $\sim$10$^{11}$ lower compared to the unsubtracted results. GPR also successfully removes the horizontal band of power at 5$\mu$s caused by the sub-band gaps.

Fig.~\ref{fig:powerspectrum-1d-after-gpr} shows the spherically-averaged power spectra that include all modes (including foreground modes) after foreground removal. Foreground removal allows the use of the low-$k$ foreground modes, down to $k=0.07$ $h$Mpc$^{-1}$. LOFAR is much more sensitive at these scales and, compared to foreground avoidance, requires less observing time to achieve comparable EoR constraints.

From the results, it is clear that GPR cannot fully remove the excess gridding power introduced by nearest neighbour gridding or insufficient sampling of the kernel, although even in those cases, it reduces the wedge power considerably. The default \textsc{wsclean} settings show an excess of 0.1 mK at low $k$ values of $0.07$~$h$Mpc$^{-1}$ up to approximately 10 mK at high $k$ values of $0.9$~$h$Mpc$^{-1}$. We define an acceptable excess power in the foreground subtraction strategy to be at most 0.1 mK at $k=0.1$~$h$Mpc$^{-1}$. With this requirement, the default settings do not result in sufficient accuracy. To reach this level of accuracy, the only parameter that requires tuning is the oversampling factor. This is in contrast to the foreground avoidance strategy, where increased $w$-quantization and oversampling factor are required to reach an acceptable level of excess power. GPR is able to remove excess power caused by $w$-quantization, making it possible to use the default of 32 $w$-layers. The GPR results with \textsc{idg} as gridding algorithm meet the required accuracy, with an excess power of 3~$\mu$K at $k$=0.1~$h$Mpc$^{-1}$ and, similar to the foreground avoidance results, overall shows the best accuracy.

\subsection{Required $w$-stacking settings}
The last column of Table~\ref{tbl:parameters} lists the minimal (least expensive) $w$-stacking gridding settings that are required to achieve a maximum excess power of 0.1~mK at $k=1$~$h$Mpc$^{-1}$ and $0.1$~$h$Mpc$^{-1}$ in the case of foreground avoidance and foreground subtraction, respectively. Compared to the default settings, constraining the excess power requires increasing the oversampling factor and the number of $w$-layers, while the kernel size and padding can be decreased.

\begin{table}
\caption{Imaging runtime on the "Dawn" cluster, using both CPU sockets of each of the 15 compute nodes. The IDG-GPU imager additionally uses one of the Tesla K40 GPUs on each node. The factor is the relative time with respect to $w$-stacking with default settings. The last column specifies the visibility throughput for a single node.}
\label{tbl:computation-requirements}
\begin{tabular}{|l|r|r|r|}
\hline
\textbf{Imaging method} & \textbf{Runtime} & \textbf{Factor} & \textbf{Throughput} \\
& & & \textbf{(KVis/s)} \Tstrut\Bstrut \\
\hline
$w$-stacking default & 4 min & 1 & 720 \Tstrut \\
($o=63$, nwl=$32$, p=1.2) &  &  &  \\
$w$-stacking minimum & 7 min & 1.75 & 410 \Tstrut \\
($o=4$K, nwl=$500$, p=1) &  &  & \\
$w$-projection & 27 min & 6.75 & 110 \Tstrut \\
Direct transform & 38 h & 570 & 1.3 \Tstrut \\
IDG-CPU & 4 min & 1 & 720 \Tstrut \\
IDG-GPU & 16 s & 0.07 & 11000 \Tstrut \\
\hline
\end{tabular}
\end{table}

\subsection{Computational requirements}
In this section we report the computational requirements for the default and minimal gridding settings as listed in Table~\ref{tbl:parameters}. We compare this to the performance of \textsc{idg} and a direct FT.  We use 15 compute nodes from the LOFAR EoR ``Dawn'' cluster, which each have the following specifications: 2 Intel Xeon E5-2670v3 CPUs (for a total of 24 physical cores), 128 GB of memory and 4 NVIDIA Tesla K40 GPUs (unless noted otherwise, we use only one GPU in our experiments). The CPUs provide a combined peak performance of 2.0 TFlop/s (single-precision, using FMA and AVX2 instructions), while one Tesla K40 GPU has a single-precision peak performance of 5.0 TFlop/s. The imaging is performed in parallel on the 15 nodes.

We measure the runtime for an imaging task that consists of creating the point spread function and the four Stokes images (I, Q, U, V) for each of the 94 sub-bands, with a total of $2.6\times 10^9$ visibilities
and report results in Table~\ref{tbl:computation-requirements}. We do not include the calculation of the LOFAR primary beam in the runtime measurement.

These results illustrate that a direct transform takes too much time in practice. $w$-projection is significantly faster (more than 84 times than the direct transform), while $w$-stacking is even faster. The difference in runtime for $w$-stacking with a larger kernel is explained as follows: (1) the number of w-layers is increased from 16 to 300 (this increases runtime); (2) the padding factor is reduced from 1.2 to 1.0 (no padding, this reduces runtime). Using kernels smaller than 7 pixels (in case of $w$-stacking or $w$-projection) does not significantly reduce runtime \citep{offringa-wsclean-2014} and we therefore use 7. The CPU version of IDG is about as fast as $w$-stacking with a kernel size of 7, while the GPU version of IDG is much faster. The accuracy of the CPU and GPU versions of IDG is the same.

\citet{veenboer-gpuidg-2017} illustrate that performance of IDG is not bound
by the number of (floating-point) operations alone. They use throughput, measured as the number of visibilities processed per
second as a (floating-point) operation-agnostic performance metric. Throughput therefore provides a meaningfull way to express imaging
performance and we will use it to compare the performance of the different imaging algorithms.

Given the imaging parameters and the runtime measurements, we compute the achieved imaging throughput per node, see the rightmost column of Table~\ref{tbl:computation-requirements}. Note that our visibility count considers the Stokes parameters separately, while \citet{veenboer-gpuidg-2017} consider the four parameters as a single visibility. Taking this into account, and correcting for the faster GPU (GeForce GTX 1080, 9.2 TFLOP/s) in their measurements, we achieve only 5\% of the throughput that they report. The difference is mainly caused by the overhead of applying the IDG gridder kernel as part of a larger application (\textsc{wsclean}) with all associated practical overheads, such as disk access and reordering of visibilities.

To put these results in perspective we also measured the calibration runtime with \textsc{sagecal-co} \citep{kazemi-2011-sagecal,yatawatta-co-2016}, which on the same compute nodes (using 15 nodes with all four GPUs) requires several days. The required imaging time is therefore not a bottleneck in the full LOFAR EoR data processing pipeline \citep{patil-2017}. Nevertheless, fast imaging is very useful for analysis.

\section{Discussion \& conclusions} \label{sec:discussion}
We have shown the bias induced by gridding visibilities on a regular grid with various settings, using traditional convolutional gridding and image domain gridding. If the brightest sources are removed before gridding, the gridding excess power resulting from traditional convolution gridding of LOFAR data sets ranges from approximately 100~mK with simple gridding settings to 10~$\mu$K with tuned gridding settings. Image domain gridding has a superior accuracy, and results without any tuning in accuracies of 2~$\mu$K at $k=0.07$~$h$Mpc$^{-1}$ in a foreground removal approach up to at most 30~$\mu$K for all measured $k$-values in both a foreground removal or foreground avoidance approach. The expected strength of the redshifted 21-cm signal is a few mK, hence the excess power caused by either gridding method can be limited to an insignificant level well below the noise level. This also shows that the SKA will not be limited by gridding noise even in extremely deep integrations. The improved $uv$-coverage of the SKA over LOFAR is likely to lower the gridding noise further.

The two parameters that are crucial for 21-cm experiments are the oversampling rate of the kernel and the quantization in the $w$-direction. The reason for this is that the discretization of $u, v$ and $w$ cause frequency-dependent errors. These spectral fluctuations make it harder to separate the astronomical foreground from the 21-cm signals. For the LOFAR EoR case, where the FOV is $3\degree$ $\times$ $3\degree$ and a maximum baseline of $250 \lambda$ is used, the kernel is required to be at least oversampled by a factor of 4000, implying a table of at least 28000 values in the case of a gridding kernel of size 7. The $w$-direction is required to have at least 500 quantization levels. Alternatively, using an algorithm without $w$-correction also produces good power spectrum results, but leads to a decorrelation loss of $\sim$8\% for the LOFAR field of view.

The current LOFAR EoR results of $\Delta^2 <$~(79.6mK)$^2$ at $k$=0.053~$h$Mpc$^{-1}$ (one night; \citealt{patil-2017}) and $\Delta^2 <$~(72.4~mK)$^2$ at $k$=0.075~$h$Mpc$^{-1}$ (10 nights; \citealt{mertens-10night-eorlimit-2019}) are not significantly affected by gridding noise. Those results use foreground subtraction and different kernel oversampling settings. In both cases a higher kernel oversampling setting was used compared to the default \textsc{wsclean} setting. The default settings would have resulted in a contribution of approximately 0.1 mK to the spherically-averaged power spectrum measurements (Fig.~\ref{fig:powerspectrum-1d-after-gpr}).

In this work, we have focussed on the imaging accuracy. A related operation that is required during calibration, is the prediction of model visibilities from a sky model. The prediction accuracy has a reciprocate relation to the imaging accuracy, and the results in this paper therefore imply that visibility prediction using gridding algorithms can be made to have sufficient accuracy for 21-cm EoR data calibration. This is crucial to calibrate on sky models with large number of sources as will be required for the SKA.

The results imply that the use of the $w$-projection algorithm \citep{wprojection-cornwell} as a $w$-term correcting algorithm is likely not an option for EoR experiments, as oversampling the gridding kernel is inherently difficult in $w$-projection due to the need for tabulating a large number of $w$-value kernels. For example, to oversample 4095 times, the memory cost for the two-dimensional $w$-kernels increases by a factor of $4095^2$. With an average kernel size of $32^2$ pixels and 512 $w$-projection planes, this would require 33 terabyte of memory. \citet{barry-2019-eppsilon} show that for a homogenous array and a beam that is separable in the direction on the sky, large oversampling is possible using \textsc{fhd}. The \textsc{idg} algorithm is an interesting alternative, in particular when ionospheric or beam terms are necessary during gridding. Faceted imaging has shown to be an effective approach for high-quality low-frequency observations \citep{kogan-greisen-2009, vanweeren-2016, tasse-2018-ddfacet}, and is for example used in the LOFAR Two-metre Sky Survey \citep{shimwell-2017-lotss}. However, faceted imaging results in discontinuities in image space, and are therefore unsuitable for 21-cm power spectra in which the Fourier modes of the image are measured.

The high accuracy and speed of \textsc{idg}, combined with its possibility for beam and ionospheric corrections, makes \textsc{idg} an attractive option for experiments that try to detect the 21-cm signals from the Epoch of Reionization. These properties will in particular be important for processing of the future Square Kilometre Array EoR observations.

\begin{acknowledgements}
We thank W. Brouw for useful comments. F.~Mertens and L.~V.~E.~Koopmans  would like to acknowledge support from a SKA-NL Roadmap grant from the Dutch ministry of OCW. S. van der Tol was supported by the Astronomy ESFRI and Research Infrastructure Cluster, part of the European Union’s Horizon 2020 research and innovation programme, under grant agreement No 653477.
\end{acknowledgements}

\label{lastpage}

\DeclareRobustCommand{\TUSSEN}[3]{#3}

\bibliographystyle{aa} 
\bibliography{references}

\end{document}